\documentclass[twocolumn]{aastex631}

\usepackage [english]{babel}
\usepackage [autostyle, english = american]{csquotes}
\MakeOuterQuote{"}
\DeclareUnicodeCharacter{2212}{-}

\begin{document}

\title{Non-detection of Helium in the Hot Jupiter WASP-48b}

\author[0000-0002-9030-0132]{Katherine A. Bennett}
\affiliation{Department of Earth and Planetary Sciences, Johns Hopkins University, Baltimore, MD 21218, USA}
\affiliation{Astronomy Department, Wesleyan University, Van Vleck Observatory, Middletown, CT, 06459, USA}

\author[0000-0003-3786-3486]{Seth Redfield}
\affiliation{Astronomy Department, Wesleyan University, Van Vleck Observatory, Middletown, CT, 06459, USA}

\author[0000-0002-9584-6476]{Antonija Oklop\v{c}i\'{c}}
\affiliation{Anton Pannekoek Institute of Astronomy, University of Amsterdam, 1098 XH Amsterdam, Netherlands}

\author[0000-0002-0810-3747]{Ilaria Carleo}
\affiliation{Instituto de Astrof\'{i}sica de Canarias (IAC), 38205 La Laguna, Tenerife, Spain}

\author[0000-0001-8720-5612]{Joe P. Ninan}
\affiliation{Department of Astronomy and Astrophysics, Tata Institute of Fundamental Research, Mumbai, 400005, India}

\author[0000-0002-7714-6310]{Michael Endl}
\affiliation{Department of Astronomy, The University of Texas at Austin, Austin, TX, 78712, USA}



\begin{abstract}

Hot Jupiters orbiting extremely close to their host star may experience atmospheric escape due to the large amounts of high-energy radiation they receive. Understanding the conditions under which this occurs is critical, as atmospheric escape is believed to be a driving factor in sculpting planetary populations. In recent years, the near-infrared 10833 \r{A} helium feature has been found to be a promising spectral signature of atmospheric escape. We use transmission spectroscopy to search for excess helium absorption in the extended atmosphere of WASP-48b, a hot Jupiter orbiting a slightly evolved, rapidly-rotating F star. The data were collected using the Habitable Zone Planet Finder spectrograph on the Hobby-Eberly Telescope. Observations were taken over the course of seven nights, from which we obtain three transits. No detectable helium absorption is seen, as absorption depth is measured to $-0.0025\pm0.0021$, or $1.2 \sigma$ from a null detection. This non-detection follows our current understanding of decreasing stellar activity (and thus high-energy radiation) with age. We use a 1D isothermal Parker wind model to compare with our observations and find our non-detection can best be explained with a low planetary mass-loss rate and high thermosphere temperature.
We explore our results within the context of the full sample of helium detections and non-detections to date. Surprisingly, comparing helium absorption with the stellar activity index $\rm log\;R'_{HK}$ reveals a large spread in the correlation between these two factors, suggesting that there are additional parameters influencing the helium absorption strength.

\end{abstract}

\keywords{Exoplanet astronomy (486) --- Exoplanet atmospheres (487) --- Hot Jupiters (753) --- High resolution spectroscopy (2096) --- Transmission spectroscopy (2133)}


\section{Introduction} \label{intro}

With the explosion of exoplanet detections in the last decade, an increasing focus has been on understanding planetary demographics and evolution. Atmospheric escape of close-in exoplanets is a phenomenon that can help us answer both of these questions. It is thought to be a partial explanation for the dearth of planets between 1.5--2 $R_{\earth}$, known as the radius valley \citep{Fulton2017}, because it may drive enough atmospheric mass loss on planets of a certain size threshold to significantly affect how a planet evolves (e.g., \citealt{Owen2013}). Additionally, quantifying mass loss rates through studies of atmospheric escape provides insight into atmospheric durability over a planet's lifetime, which in turn is important for understanding habitability \citep{Lammer2014}. 

Thought to be driven primarily by high-energy ultraviolet (UV) and X-ray stellar radiation (e.g., \citealt{Lammer2003, Murray-Clay2009}), atmospheric escape was first directly observed in the atmosphere of HD 209458b \citep{Vidal-Madjar2003} via the detection of the Ly-$\alpha$ hydrogen transition in the planetary atmosphere using UV transmission spectroscopy. However, to date only a handful of detections using this marker have been made (e.g, \citealt{Lecavelier2012, Ehrenreich2012, Kulow2014}). Though the line has a large cross section \citep{Owen2019}, the lack of ease of detection is attributable to the fact that the line is in the ultraviolet regime (at 1215 \r{A}) and thus cannot be observed using high-resolution ground-based facilities. In addition, the core of the line is altered by heavy absorption by the interstellar medium (ISM) \citep{Wood2005, Edelman2019} and Earth's own geocoronal emission (e.g., \citealt{Vidal-Madjar2003}). 

In addition to Ly-$\alpha$, the H$\alpha$ line of the Balmer series has also been observed in a handful of exoplanetary atmospheres (e.g., \citealt{Jensen2012, Yan2018}).  However, there have been controversies over whether some detections of H$\alpha$ are due to stellar activity or planetary absorption (e.g., \citealt{Barnes2016, Cauley2017}). 

Metals, including OI \citep{Vidal-Madjar2004, Ben-Jaffel2013}, CII \citep{Vidal-Madjar2004, Linsky2010}, MgI \citep{Vidal-Madjar2013}, and MgII and FeII \citep{Sing2019} have also been used as indicators of atmospheric escape. Additionally, broadband X-ray transit observations of HD~189733b have shown a discrepancy between the X-ray and optical transit depths, supporting the existence of escaping metals (which absorb at X-ray wavelengths) in planetary exospheres \citep{Poppenhaeger2013}. 

The metastable helium feature at 10833 \r{A} has also emerged in recent years as a marker of atmospheric escape. It is less susceptible to stellar activity than H$\alpha$ and is not affected by absorption from the ISM like Ly-$\alpha$. In addition, a huge advantage of using this feature over Ly-$\alpha$ is that is in the infrared regime, meaning it can be observed using ground-based telescopes. 

While it was first predicted to be a strong absorption feature of transmission spectroscopy in 2000 \citep{Seager2000}, this triplet feature was not detected in an exoplanetary atmosphere until 2018, when it was observed in the atmosphere of WASP-107b using narrowband photometry with the Wide Field Camera 3 (WFC3) onboard the \textit{Hubble Space Telescope (HST)} \citep{Spake2018}. Since then, the number of helium detections has exploded. To date, more than 30 planets have been investigated for the 10833 \r{A} helium line. Of these, most have been nondetections (e.g., \citealt{Kasper2020, Krishnamurthy2021, Vissapragada2021, Zhang2021, Kawauchi2022}), but there have been many detections as well (e.g., \citealt{Nortmann2018, Allart2018, Salz2018, Ninan2020, Paragas2021}). 

The helium absorption feature at 10833 \r{A} is due to resonance scattering of neutral helium from the $\rm 2^3S$ to the $\rm 2^3P$ state \citep{Oklopcic2018}. There are two main pathways to populate the $\rm 2^3S$ metastable state: photoionization-recombination and collisional excitation \citep{Oklopcic2018}. In the photoionization-recombination pathway, EUV photons are responsible for ionizing neutral helium, which then recombines to the metastable $\rm 2^3S$ state. In exoplanets, this is the dominant pathway populating the metastable helium state. Therefore, the population of this  state (and thus helium absorption at 10833 \r{A}) is thought to be driven primarily by the amount of EUV flux received from the host star \citep{Oklopcic2019}. Accordingly, an intense amount of focus has been placed on determining a correlation between EUV flux and helium absorption depth.

On the other hand, the metastable $\rm 2^3S$ state is ionized and depopulated by mid-UV flux \citep{Oklopcic2019}. This suggests larger helium absorption features may be expected from planets orbiting stars with high amounts of EUV radiation, which originates from the stellar corona \citep{Nortmann2018}, but low amounts of mid-UV blackbody radiation. The optimal ratio of EUV to mid-UV flux was found by \cite{Oklopcic2019} to exist around K-type stars, and this finding has largely skewed the sample of helium studies toward K-type stars. 

Additionally, because EUV flux originates from stellar magnetic activity in the corona, and stars are most active in their younger years \citep{Ribas2005}, emphasis has also been placed on helium searches for planets orbiting young, active stars. While it is important to continue to study atmospheric escape in these types of environments, in order to fully comprehend the relationship between high-energy radiation and helium absorption, environments other than planets orbiting young, K-type stars must be considered. Only with a more complete sample can we begin to sketch the story of how atmospheric escape drives planetary evolution across all stellar types and ages. 

In this paper, we use transmission spectroscopy to examine the 10833 \r{A} line in the atmosphere of a hot Jupiter orbiting a slightly evolved star. WASP-48 is an F-type star with an age of approximately 7.9 Gyr \citep{Enoch2011}, and its sole known planet is a hot Jupiter with a period of 2.14 days. Stellar and planetary parameters are given in Table \ref{wasp48_params}. This system is very similar to that of HAT-P-32, which has one of the strongest helium signals detected to date \citep{Czesla2022}. Though the HAT-P-32 system is likely younger than the WASP-48 system, both consist of hot Jupiters with similar planet properties orbiting an F-star at approximately 0.034 au (see Table \ref{stellar_params}).  

By investigating a non K-type star that is off the main sequence, we hope to contribute to a more complete sample of helium observations, which is important in order to uncover the full picture of how EUV flux impacts atmospheric escape over the course of a planet's lifetime. To the best of our knowledge, this is only the second time a star off the main sequence has been targeted in the metastable helium search. The first was WASP-12b \citep{Kriedberg2018}, in which helium was not detected. 

While the host star has a low stellar activity index (log~$\rm R'_{HK}=-5.135$), it is rotating at $v\,{\rm sin}\,i\sim12.2\;{\rm km\,s^{-1}}$, a rapid rate for a star of its age. This make it an intriguing candidate for a metastable helium study, as a higher rotation rate is suggestive of greater stellar activity. Therefore, a search for helium in this exoplanet will provide insight into whether evolved stars can indeed show high levels of EUV flux and thus drive atmospheric escape beyond the time frame accepted in our current models.
\begin{deluxetable}{cc}
\tablecaption{Parameters of WASP-48b and its host star \label{wasp48_params}}
\tablenum{1}
\tablewidth{0pt}
\def\arraystretch{.85}
\tablehead{\colhead{Parameter} & \colhead{Value}}
\startdata
         Distance (pc) & 460 ± 5 \\
         Stellar Radius ($R_{\odot}$) & 1.75 ± 0.09 \\ 
         Stellar Mass ($M_{\odot}$) & 1.19 ± 0.05 \\
         Stellar $T_{\rm eff}\,$ (K) & 5920 ± 150 \\
         Stellar Age (Gyr) & $7.9^{+2.0}_{-1.6}$ \\
        Spectral Type & F\\
        V magnitude & 11.72 ± 0.14 \\
        log $g_{\odot}$ $(\rm cm\;s^{-2}$) & 4.03 ± 0.04 \\
         $\rm [Fe/H]$ & $-0.12\pm0.12$\\
         $v\,{\rm sin}\,i\;({\rm km\,s^{-1}})$ & 12.2 ± 0.7 \\
         Planet Radius ($R_{J}$) & 1.67 ± 0.10 \\ 
         Planet Mass ($M_{J}$) & 0.984 ± 0.085 \\
         Planet Density ($\rho_J$) & 0.21 ± 0.04 \\
         Planetary $T_{\rm eff}$ (K) & 2035 ± 52 \\
         $a$ (au) & 0.0344 ± 0.0026\\
         $P$ (days) & 2.14363592 ± 0.0000046 \\
         log $g_P$ $(\rm cm\;s^{-2}$) & 2.91 ± 0.06 \\
         Transit Duration (min) & 192.20 ± 1.73 \\
         $T_{c}(0)\;(\rm BJD_{TDB})$ & 2455364.55217 ± 0.00020 \\
         Inclination (degrees) & 80.09 ± 0.55\\
\enddata
\tablecomments{Values taken from \cite{Enoch2011} and \cite{Turner2016}.}
\end{deluxetable}

We detail the observations and analysis in Section \ref{observations}. In Section \ref{results}, we examine how our observations compare to the model derived by \cite{Oklopcic2018} and perform a more robust statistical examination of the result. We examine our findings in the context of the complete sample of helium literature in Section \ref{discussion} and work toward uncovering a relationship between helium absorption and measures of stellar activity.

\section{Observations and Data Analysis} \label{observations}

\subsection{Observations and Data Reduction}

Observations were taken using the Habitable-Zone Planet Finder (HPF), a fiber-fed NIR echelle spectrograph on the 10 m Hobby-Eberly Telescope (HET). Though HPF was originally designed to detect low-mass planets around M dwarfs using precision radial velocity techniques  \citep{Mahadevan2012}, it has become widely used in the search for helium in exoplanets (e.g., \citealt{Ninan2020, Krishnamurthy2021, Vissapragada2021}). HPF has a wavelength range coverage of 8079--12786 \r{A} and resolving power of R$\sim$55,000 \citep{Ninan2020}. 

Observations were taken on seven nights between 2019 May 16 UT and 2019 June 29 UT. These are detailed in Table \ref{obs_table}. Due to the design of the HET, observations can only take place over a limited time span, so observing a full transit from ingress to egress in one night is not possible. Instead, observations are taken on multiple nights to piece together as much of the transit as possible. Across these seven nights of observations, three transits were observed: on 2019 May 19 UT, 2019 June 16 UT, and 2019 June 18 UT. 
\begin{deluxetable}{cccc}
\tablecaption{HPF Spectrographic Observations of WASP-48b.
\label{obs_table}}
\tablenum{2}
\tablewidth{0pt}
\def\arraystretch{.85}
\tablehead{\colhead{Date} & \colhead{Time} & \colhead{In vs. Out} & \colhead{Average} \\
\colhead{(UT)} & \colhead{(UT)} & \colhead{of Transit} & \colhead{SNR}}\
\startdata
        2019 May 16 & 08:06:44 & out & 112.3 \\
        2019 May 16 & 08:22:33 & out & 112.9\\
        2019 May 16 & 08:38:21 & out & 105.4\\
        2019 May 17 & 07:55:23 & out & 66.2\\
        2019 May 17 & 08:11:11 & out & 84.0\\
        2019 May 17 & 08:27:00 & out & 90.0\\
        2019 May 19 & 07:21:18 & out & 92.8\\
        2019 May 19 & 07:37:07 & in & 110.2\\
        2019 May 19 & 07:52:56 & in & 121.8\\
        2019 May 19 & 08:08:45 & in & 120.5\\
        2019 May 19 & 08:24:33 & in & 108.1\\
        2019 May 19 & 08:40:22 & in & 105.0\\
        2019 Jun 16 & 05:44:02 & in & 59.1\\
        2019 Jun 16 & 05:59:50 & in & 61.1\\
        2019 Jun 16 & 06:15:39 & in & 98.3\\
        2019 Jun 16 & 06:31:28 & in & 92.0\\
        2019 Jun 16 & 06:47:17 & in & 88.0\\
        2019 Jun 16 & 07:04:31 & in & 74.1\\
        2019 Jun 18 & 05:56:08 & out & 94.3\\
        2019 Jun 18 & 06:11:57 & out & 91.9\\
        2019 Jun 18 & 06:27:46 & out & 79.8\\
        2019 Jun 18 & 06:43:34 & out & 73.4\\
        2019 Jun 18 & 10:20:41 & in & 78.6\\
        2019 Jun 18 & 10:36:30 & in & 92.2\\
        2019 Jun 18 & 10:52:19 & in & 31.7\\
        2019 Jun 18 & 11:08:08 & out & 49.5\\
        2019 Jun 19 & 05:38:12 & out & 62.2\\
        2019 Jun 19 & 05:54:01 & out & 75.6\\
        2019 Jun 19 & 06:09:50 & out & 82.2\\
        2019 Jun 19 & 06:25:39 & out & 76.2\\
        2019 Jun 29 & 09:56:16 & out & 105.1\\
        2019 Jun 29 & 10:12:05 & out & 98.2\\
        2019 Jun 29 & 10:27:54 & out & 90.1\\
        2019 Jun 29 & 10:43:43 & out & 86.6\\
\enddata
\end{deluxetable}

\begin{figure}[h]
    \centering
    \hspace*{-1cm}
    \scalebox{0.52}{\includegraphics{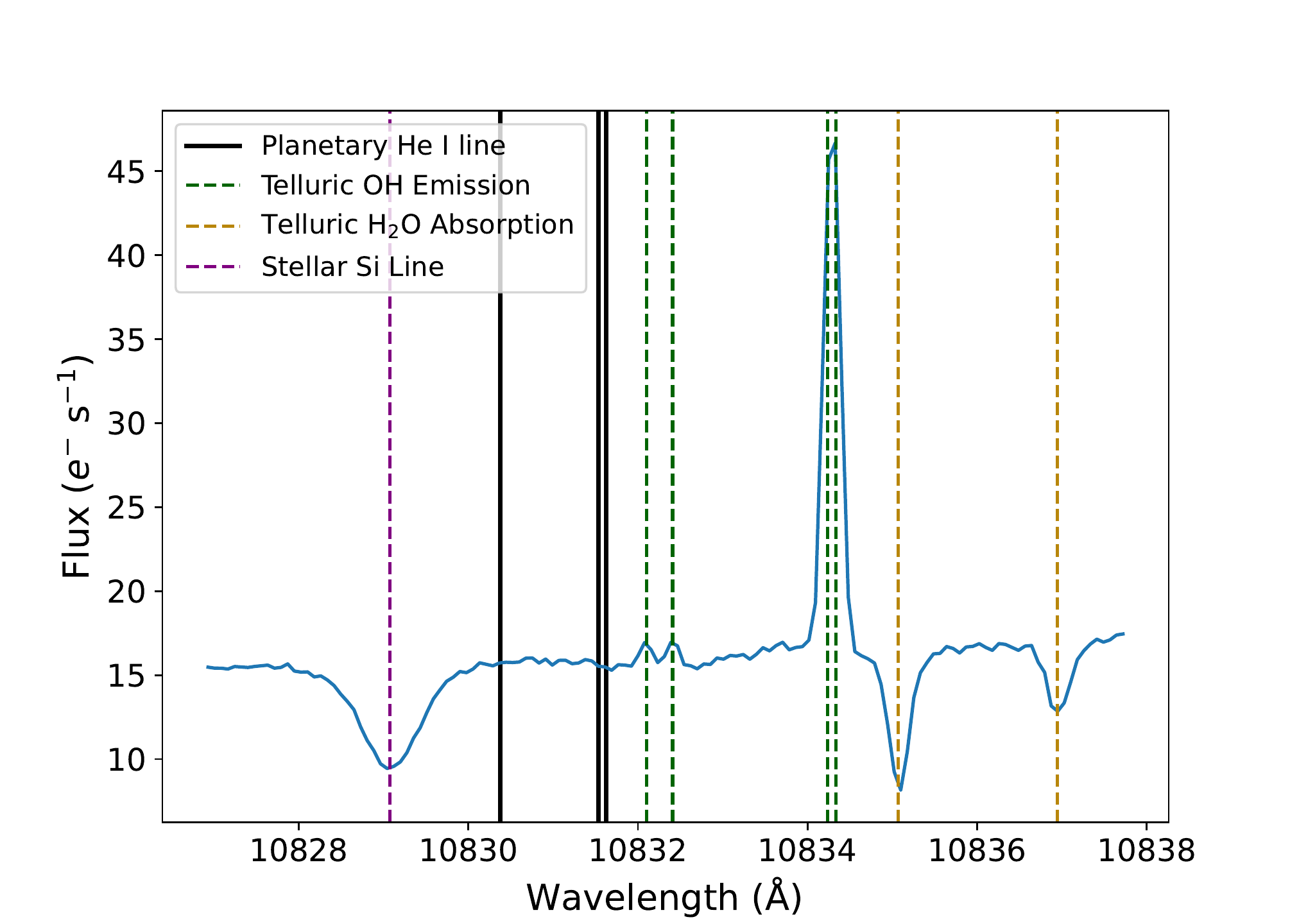}}
    \caption{Example spectrum from 2019 May 19 08:08 UT in the topocentric rest frame. Rest wavelengths of HITRAN derived OH and $\rm H_2O$ features are shown with vertical green and yellow dashed lines, respectively. These coincide with the emission lines seen in the spectrum, confirming they are telluric features originating in the Earth's atmosphere. The expected location of the stellar silicon feature and any planetary helium feature is also plotted. Note that the location of the predicted helium feature varies across exposures due to changing barycentric and planetary velocity corrections.}
    \label{single_spectrum}
\end{figure}

In our analysis, we include exposures with an average SNR of greater than 50, which excludes the exposures from 2019 Jun 18 10:52 UT and 2019 Jun 18 11:08 UT. This decision was supported by the fact that a strong stellar silicon feature occurring just blue-ward of the helium triplet (see Figure \ref{single_spectrum}) was absent in these two exposures but present in all other exposures, suggesting that WASP-48 was not successfully observed in these two cases. 

Two more exposures, those from 2019 Jun 16 05:44 UT and 2019 Jun 16 05:59 UT were also eventually discarded due to an imperfect data analysis process that proved to be particularly errant for these two exposures. More details on this realization and justification are provided in Section \ref{results}.

The data were reduced using the HPF extraction pipeline \citep{Ninan2018}. After accounting for bias removal, nonlinearity correction, and cosmic ray correction, this algorithm nondestructively measures the rate of charge accumulation per pixel in an "up-the-ramp" fashion. These 3D data are then collapsed to a 2D flux image, which is flat-fielded and rectified \citep{Kaplan2019}. Finally, an optimal extraction algorithm \citep{Horne1986} is used to weight the data by its uncertainty and cross-dispersion profile and collapse it into a 1D spectrum.

Each resulting exposure contains a science, sky, and calibration fiber spectrum, with all wavelengths being measured in vacuum. The metastable helium feature is actually a triplet with rest vacuum wavelengths of 10832.07, 10833.22, and 10833.31 \r{A}. We calculate these values by converting the air wavelengths of the helium triplet, given on the NIST Atomic Spectra Database\footnote{\url{https://www.nist.gov/pml/atomic-spectra-database}}, into vacuum wavelengths. A first glimpse of the spectrum around 10833 \r{A} can be seen in Figure \ref{single_spectrum}. A stellar silicon line and several telluric features are visible. These telluric features include an OH emission doublet at 10832.10 and 10832.41 \r{A}, another strong but blended OH emission doublet at 10834.24 and 10834.33 \r{A}, and two water vapor absorption lines at 10835.07 and 10836.94 \r{A}. The identities of these features were uncovered using the HITRAN (high-resolution transmission molecular absorption) database\footnote{\url{https://hitran.org}}. 

\subsection{Analysis} \label{analysis}

Telluric lines pepper our observations and must be removed before preceding. In particular, we focus on the OH telluric emission lines closest to the 10833 $\rm \AA$ line. When shifting the spectra into a stellar rest frame, the blue-most OH emission doublet overlays the redward components of the helium triplet (10833.22 and 10833.31 \r{A}), which will interfere with our ability to detect any planetary helium in exposures close to mid-transit, where the planetary velocity shifts are relatively small.

\begin{figure}[h]
    \centering
    \hspace*{-0.9cm}
    \scalebox{1.3}{\includegraphics[scale=0.41]{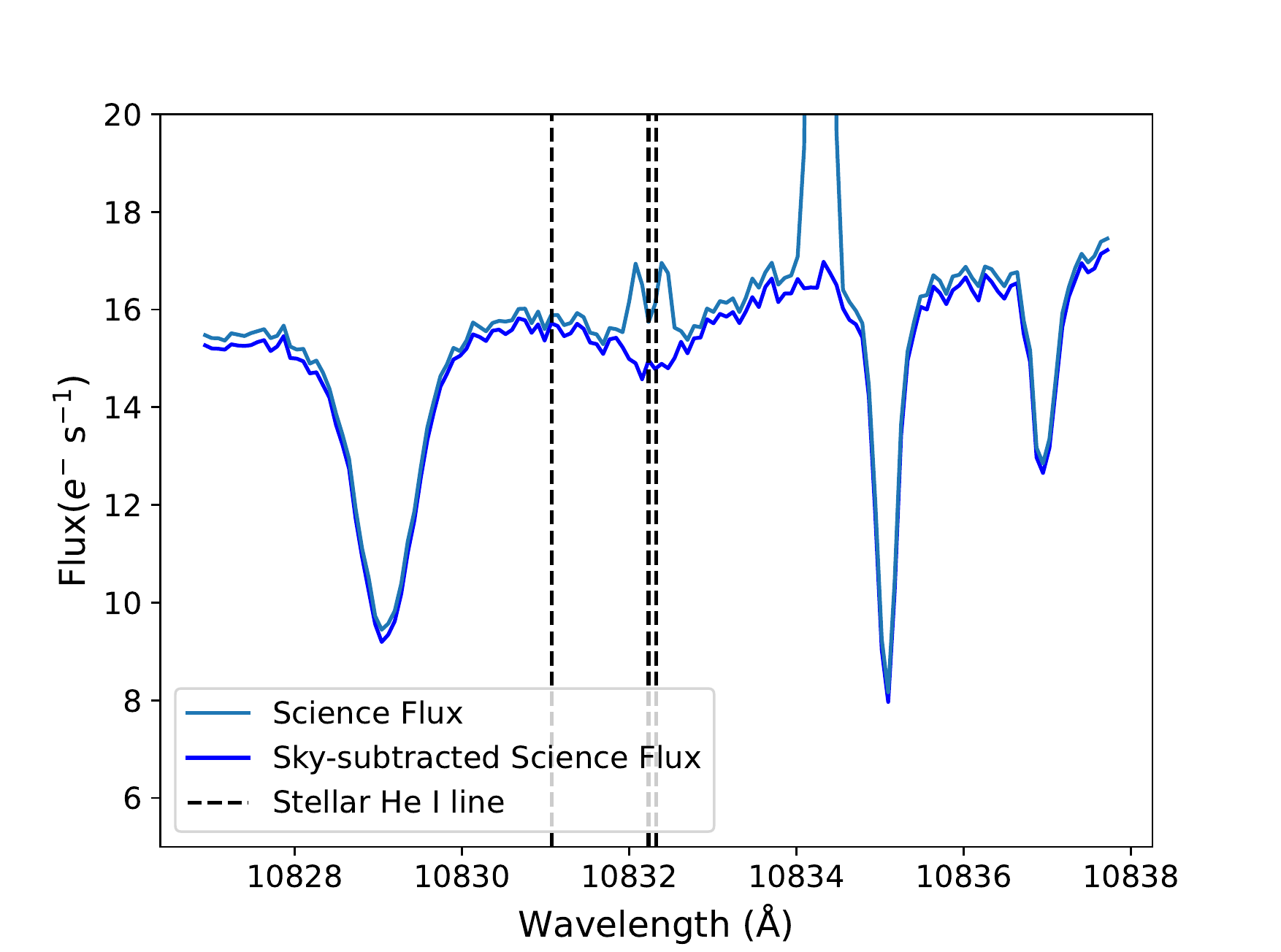}}
    \caption{Science fiber in the topocentric rest frame, both before (light blue line) and following (dark blue line) sky flux subtraction. This exposure is from 2019 May 19 08:08 UT. We have zoomed in to see the sky subtraction has successfully removed the OH emission lines and has revealed a stellar helium line underneath the small OH doublet.}
    \label{sky_sub}
\end{figure}

To remove these telluric emission lines, we subtract the sky fiber from the science fiber. For each spectra, we first remove the sky flux continuum and interpolate the sky fiber to the science fiber. We calculate a scaling ratio for the sky spectrum to correct for the varying intensity of the OH peaks between the sky and science fibers. We apply this scaling ratio to the sky spectrum between 10831--10835 $\rm \AA$. Though the scaling ratio does have a chromatic dependence, the two OH peaks are very close in wavelength, and one scaling factor should suffice. Once the scaling factor is applied, we subtract the interpolated sky spectrum from the science spectrum. This technique reveals a shallow stellar helium feature underneath the OH emission lines, as shown in Figure \ref{sky_sub}.  

By examining Figure \ref{sky_sub}, we can see that the closest telluric water absorption lines falls about 2 \r{A} from the stellar helium feature. We calculate that the maximum planetary radial velocity in-transit is $\rm 32.7\;km\;s^{-1}$, which translates into a $\rm 1.2\;\AA$ shift at 10833 \r{A}. Because this is smaller than the wavelength difference between helium and the telluric water lines, we are justified in disregarding those telluric absorption lines. (We also conducted a visual examination of each in-transit spectrum in the planet rest frame to confirm that the expected planetary helium feature location does not overlap in any instance with the telluric water lines.)

Once the telluric features have been accounted for, we create a master out-of-transit spectrum. First, the spectra are normalized by dividing each science spectra by a third-order polynomial fitted to the stellar continuum, following \cite{Salz2018}. Next, each out-of-transit spectrum is shifted from a topocentric rest frame into the stellar rest frame. To do this, the spectra are first shifted into a barycentric rest frame using the \texttt{barycorrpy} Python package developed by \cite{Kanodia2018}, which is based on the IDL code by \cite{Wright2014}. We then immediately shift each spectrum into WASP-48's stellar rest frame by accounting for the systemic radial velocity of WASP-48 ($-19.740$ $\rm km$ $\rm s^{-1}$, taken from the SIMBAD Astronomical Database\footnote{\url{http://simbad.u-strasbg.fr/simbad/}}). We ignore the reflex stellar radial velocity from the motion of the planet, because we calculate the radial velocity semiamplitude of the star to be only $0.14$~$\rm km$ $\rm s^{-1}$. At 10833 $\rm \AA$, this translates to a shift of $\rm 0.0050\;\AA$, which is less than the spectral resolution of HPF at this wavelength ($\rm \sim0.2\;\AA$). 

Next, we interpolate each individual out-of-transit spectrum to a single reference wavelength grid using the \texttt{scipy} cubic interpolation function. We sum the spectra together, with each spectrum weighted by the inverse of its variance following \cite{Salz2018} and \cite{Ninan2020}, to arrive at our master out-of-transit spectrum. 

\begin{figure}[h]
    \centering
    \hspace*{-1.0cm}
    {\includegraphics[scale=0.47]{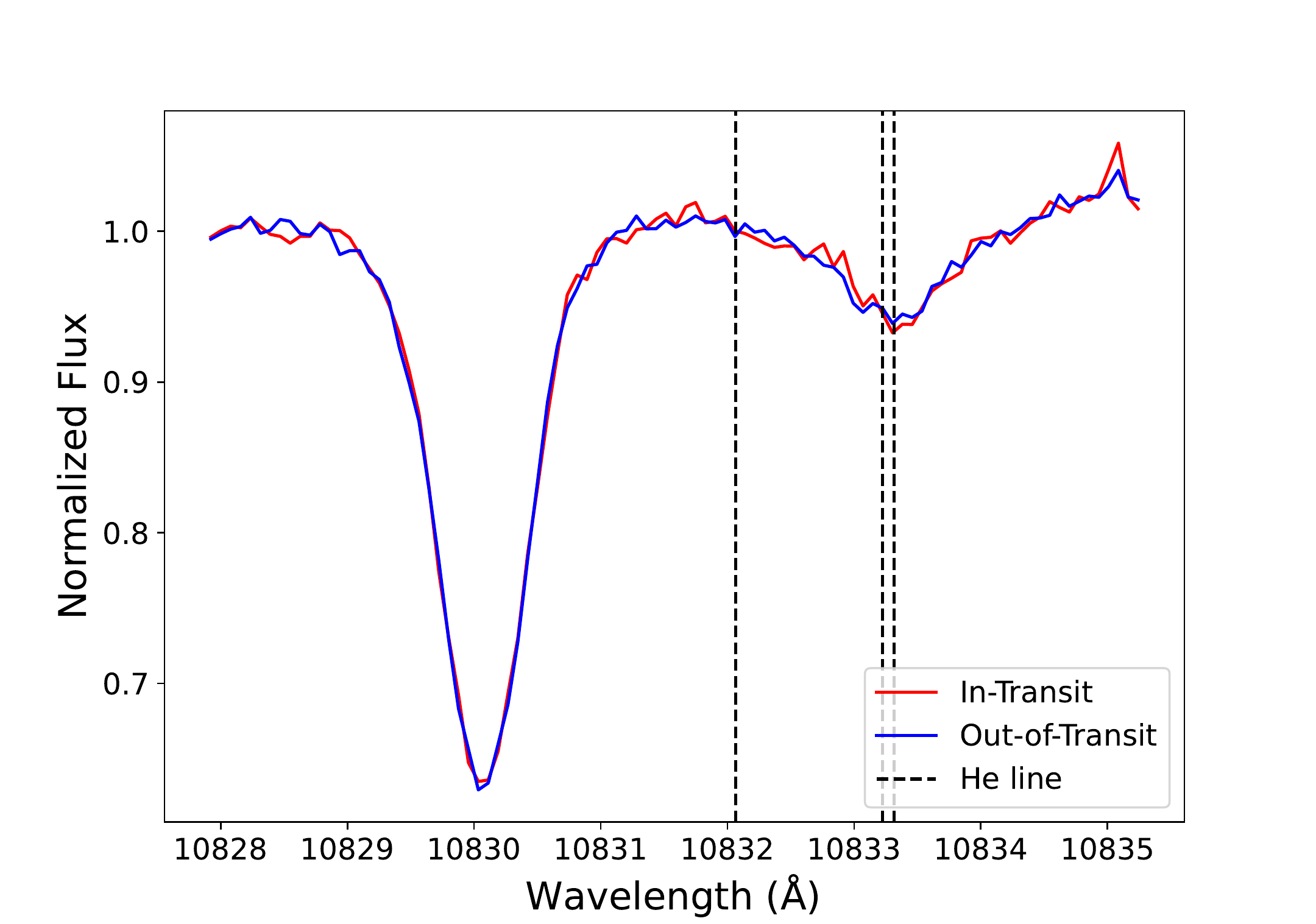}}
    \caption{In-versus-out of transit spectrum in the stellar rest frame. The individual spectra have been interpolated and undergone a weighted sum with their inverse variances used as weights. No planetary helium signal is visible.}
    \label{in out}
\end{figure}

Before shifting the in-transit spectra into their appropriate planetary rest frames, we can create an analogous master in-transit spectrum in the stellar rest frame to compare with the out-of-transit spectrum. This provides a quick visual inspection of the helium feature to see if there is any hint of planetary helium. As shown in Figure \ref{in out}, no visible planetary helium is present when examining the master in-versus-out spectrum. 

\begin{figure*}
    \centering
    \scalebox{0.5}{\includegraphics{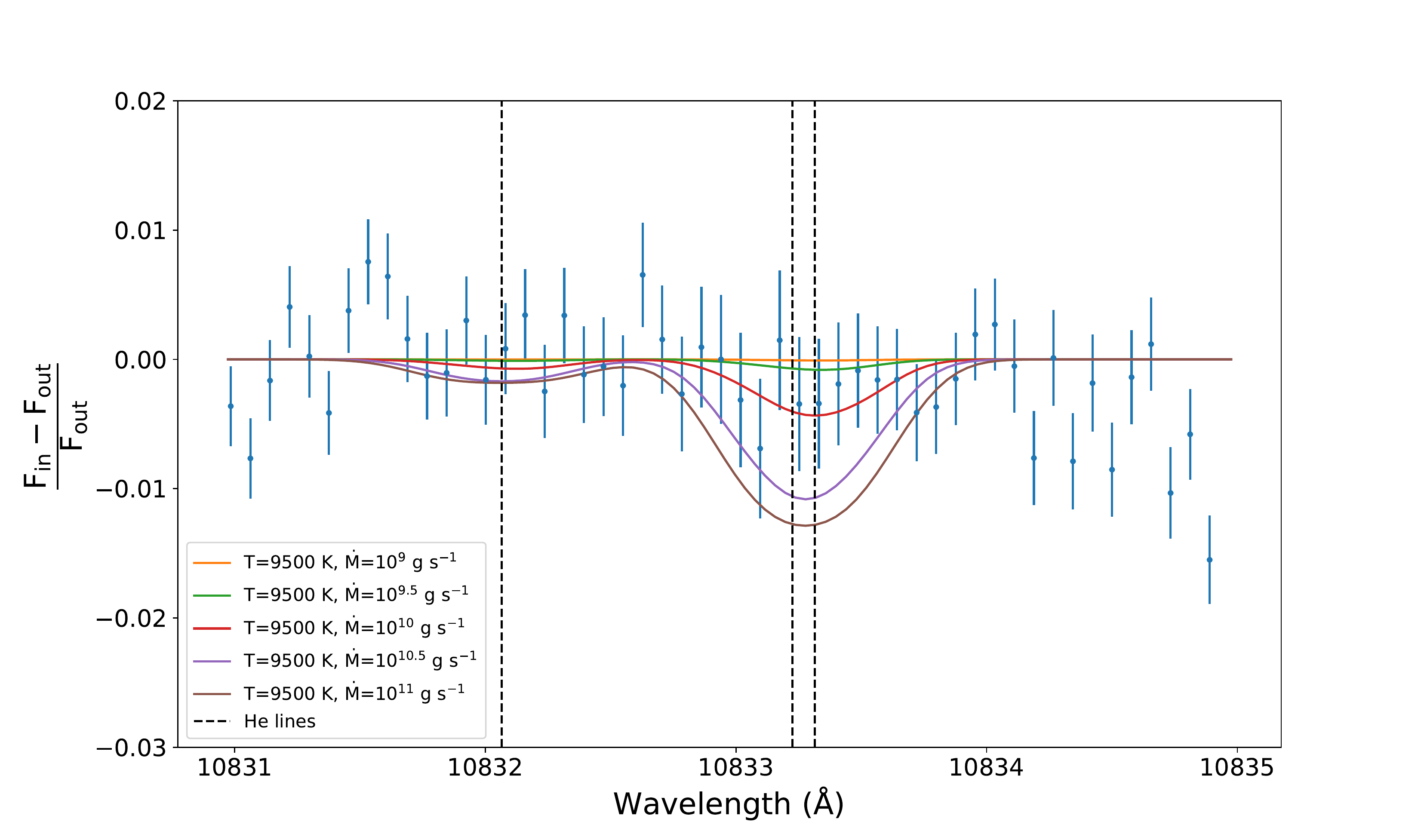}}
    \caption{Final transmission spectrum in the planetary rest frame centered on the 10831--10835 $\rm\AA$ range. Planetary absorption would be seen as a negative value, so this flat spectrum shows no indication of helium absorption. One example of the 1D Parker wind model is overplotted with the spectrum. This example depicts various mass-loss rates ($\rm \dot{M}=10^9-10^{11}\;g\;s^{-1}$) with a thermosphere temperature of 9500 K. Absorption signature increases with mass-loss rate at this thermosphere temperature.}  
    \label{model_compare_ex}
\end{figure*}

\begin{figure*}
    \centering
    \scalebox{0.48}{\includegraphics{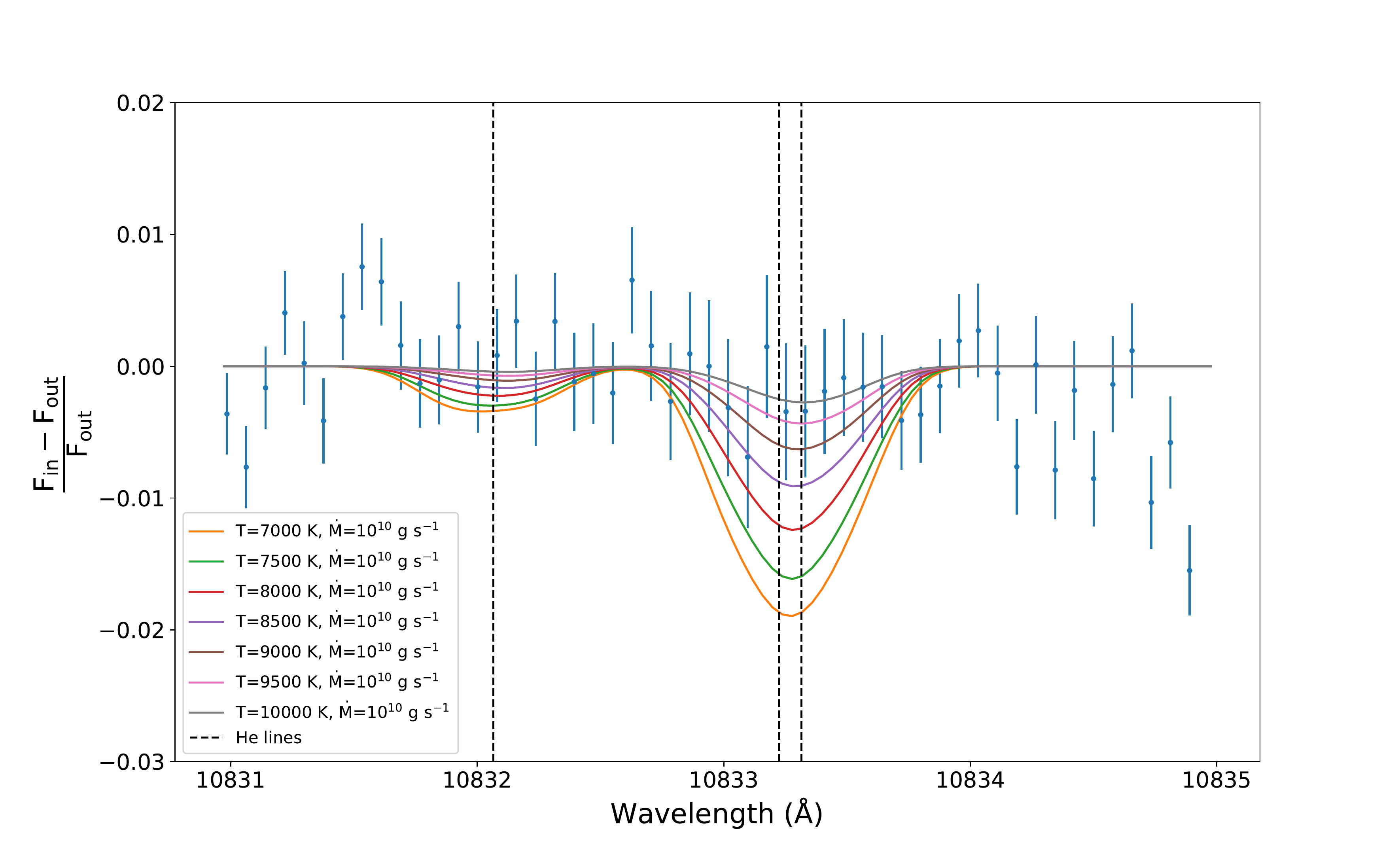}}
    \caption{Final transmission spectrum in the planetary rest frame. Same as Figure \ref{model_compare_ex}, but with varying thermosphere temperatures ($\rm T=7000-10000\;K$) at a constant mass loss rate of $\rm \dot{M}=10^{10}\;g\;s^{-1}$. At this mass loss rate, helium absorption increases with decreasing temperature.}  
    \label{model_compare_ex_temps}
\end{figure*}

To create the transmission spectrum, we interpolate the master out-of-transit spectrum to each individual in-transit spectrum and calculate: 

\begin{equation}
    \frac{F_{\rm in}-F_{\rm out}}{F_{\rm out}}=\frac{F_{\rm in}}{F_{\rm out}}-1,
    \label{trans_spec_eq}
\end{equation}

\noindent to create a series of "ratio spectra". We then shift these spectra into their planetary rest frames by using the \texttt{radvel} package \citep{Fulton2018} to calculate planetary radial velocity for each in-transit exposure, which is then converted into a wavelength shift. We interpolate the ratio spectra to a single reference grid and sum them using their inverse variances as weights. This brings us to the final transmission spectrum, shown in Figures \ref{model_compare_ex} and \ref{model_compare_ex_temps}. The spectrum is quite flat, demonstrating that there is no obvious planetary helium absorption in the atmosphere of WASP-48b. We additionally examine the transmission spectra per night and find they are also flat. 

\section{Results} \label{results}

In the formulation in Equation \ref{trans_spec_eq}, planetary absorption is expressed as as a negative value, emission as a positive value, while a value of zero indicates no planetary absorption or emission. To calculate the average absorption depth, we take the average of the absorption depth over a narrow range of wavelengths (10833.10--10833.49 $ \rm \AA$) centered on the two red components of the helium triplet, following \cite{Alonso2019} and \cite{Kirk2020}. Because the blue-most helium line (at 10832.07 $\rm \AA$) is much weaker than the redder two lines (e.g., \citealt{Oklopcic2018}), we exclude this portion of the wavelength range. We calculate the the average absorption to be $-0.0025\pm0.0021$ with a significance of $1.2 \sigma$. Therefore, this is a non-detection of metastable helium in the atmosphere of WASP-48b.

\subsection{Empirical Monte Carlo} \label{emc}

To examine the impact individual exposures taken over the course of many nights have on the final transmission spectrum, we turn to the empirical Monte Carlo (EMC) diagnostic \citep{Redfield2008}. This is a technique that aims to assess any underlying systematic errors in our analysis, including those arising from the data analysis itself as well as from astrophysical sources including stellar variability. 

The EMC applies our data analysis pipeline to a random subset of observations to arrive at a final absorption depth. It then conducts this random selection many times. Assuming no subset of exposures is skewing our overall result, the resulting histogram of absorption values should be centered around the absorption depth of  $-0.0025\pm0.0021$. To preserve the integrity of our analysis, a random selection of five in-transit and eight out-of-transit exposures is selected for each iteration, so that the ratio of in-transit to out-of-transit exposures in the EMC most closely matches the true ratio of all the in-transit exposures (11) to out-of-transit exposures (19). 

In addition to conducting the EMC for in-transit versus out-of-transit exposures ("in-out"), we also apply the EMC to compare a random subset of in-transit exposures with the rest of the in-transit exposures ("in-in"). We do the same for out-of-transit exposures ("out-out"). These comparisons are done to ensure there is no systematic deviation within the in-transit or out-of-transit data. The absorption depth for both of these comparisons should be zero, signifying that there is no difference between various in-transit exposures (and the same for out-of-transit exposures). As in the "in-out" comparison, we preserve the in/out ratio for these two cases as closely as possible. 

The EMC was run for 330 iterations in all three scenarios: "in-out", "out-out", and "in-in". 330 iterations is the maximal possible number of combinations for the "in-in" scenario and thus became the limiting factor. 

At this stage, two more exposures were excluded in the analysis: 2019 Jun 16 05:44 UT and 2019 Jun 16 05:59 UT. When running the EMC with these two exposures included, instead of a Gaussian shape centered around the calculated absorption value, there is a clear bimodal distribution in the "in-out" scenarios. This indicates that a small subset of exposures is largely skewing the data. When these two exposures are removed, the bimodal nature of the "in-out" distribution disappears. These two exposures also had the lowest SNR of our sample and were causing errant emission in our transmission spectrum.

The final EMC is seen in Figure \ref{emc}, and the calculated absorption and standard deviation for the three scenarios are shown in Table \ref{emc_info}. From Figure \ref{emc}, though the distributions in the three scenarios do not perfectly align, there is no indication that the "in-out" data are significantly different from the "in-in" or "out-out" scenarios. This is supported by Table \ref{emc_info}, which demonstrates that all three scenarios are within 1$\sigma$ of zero absorption (and within 2$\sigma$ of each other), further supporting our conclusion. We can also compare the calculated absorption and standard deviation from the transmission spectrum (black lines in Figure \ref{emc}) to the mean and standard deviation of the "in-out" scenario of the EMC (magenta lines in Figure \ref{emc}; see also Table \ref{emc_info}). From the transmission spectrum, we measure absorption to be $-0.0025\pm0.0021$, and in the EMC, we measure $-0.0023\pm0.0057$. This is within $0.1\sigma$ of the calculated absorption value, and so we can confirm that we sufficiently accounted for systematic errors in our analysis. 

\begin{figure*}
    \centering
    \hspace*{-0.75cm}
    \scalebox{0.58}
    {\includegraphics{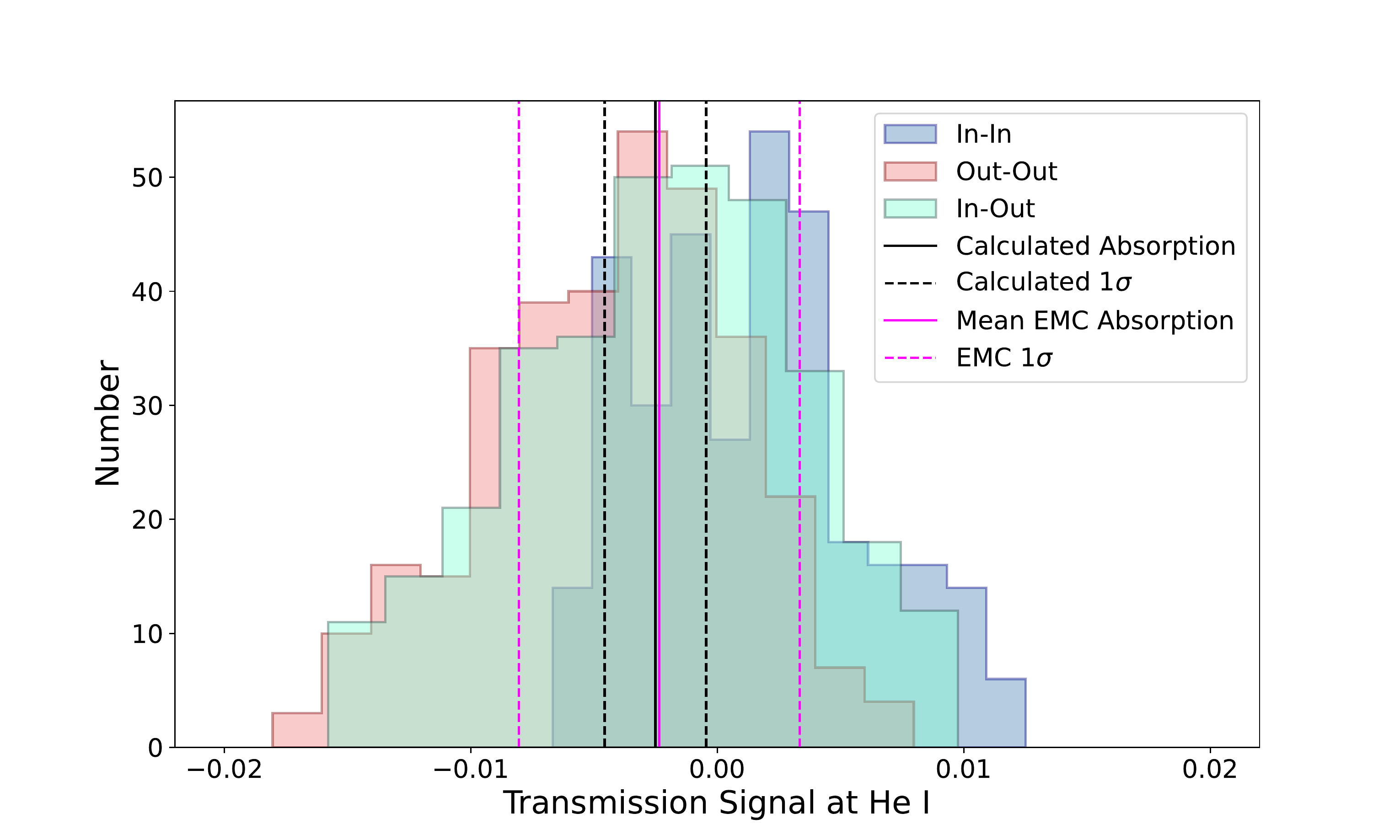}}
    \caption{Final version of EMC, showing absorption depth at the 10833 $\rm \AA$ helium line across randomly selected subsets of in and out-of-transit exposures. The solid and dotted black lines depict the calculated absorption depth and error from the transmission spectrum. The corresponding magenta lines show the average absorption and standard deviation calculated in the EMC. The lines are within $0.1\sigma$ of each other, suggesting we have correctly accounted for systematic errors in the analysis.}
    \label{emc}
\end{figure*}

\begin{deluxetable}{lll}
\tablecaption{EMC distribution properties.
\label{emc_info}}
\tablenum{3}
\tablewidth{200pt}
\tablehead{\colhead{} & \colhead{$\mu$} & \colhead{$\sigma$}}
\startdata
         In-In & 0.0014 & 0.0044\\ 
         Out-Out & $-0.0043$ & 0.0050\\
         In-Out & $-0.0023$ & 0.0057\\
\enddata
\end{deluxetable}

\subsection{Modeling} \label{model}

We use the 1D isothermal Parker wind model developed by \cite{Oklopcic2018} to constrain the mass loss rate of WASP-48b based on helium absorption. Briefly, the model uses input parameters of the stellar spectrum, planetary mass and radius, planetary atmospheric composition, temperature of the thermosphere, and mass-loss rate to derive density and velocity profiles of the atmosphere as a function of altitude. These profiles are used to determine the population levels of helium atoms as a function of radius, which are then used to ultimately calculate the column density of the metastable $\rm 2^3S$ state as a function of radius. From the column density profile, the predicted metastable helium absorption is calculated by integrating the optical depth from the planetary radius to the planetary Hill radius.

Input parameters of planetary mass and radius are given in Table \ref{wasp48_params}. Solar composition (9:1 hyrogen-to-helium ratio) and solar input spectrum are assumed. As described in \cite{Oklopcic2019}, the solar spectrum is derived from the Solar Radiation and Climate Experiment (SORCE) solar spectral irradiance data from the Laboratory for Atmospheric and Space Physics (LASP) Interactive Solar Irradiance Data Center\footnote{\url{http://lasp.colorado.edu/lisird/}}. This spectrum includes the region 5--395 \r{A} and $\sim$1150--23700 \r{A}. To account for the EUV irradiance between $\sim$400--1150 \r{A}, the Ly$\alpha$ scaling relation described in \cite{Linsky2014} is applied. The complete input solar spectrum for this work can be viewed as the G2 model in Figure 1 of \cite{Oklopcic2019}. 

Unfortunately, it is difficult to a priori estimate the planetary thermosphere temperature and mass loss rate without information about the the high-energy spectrum of the star. The XUV spectrum of WASP-48b is virtually unknown, but we can estimate a range of predicted mass-loss rates using the energy-limited formula (e.g., \citealt{Murray-Clay2009}):
\begin{equation}
    \dot{M}\approx\frac{\epsilon \pi R^3_P F_{XUV}}{G M_P}.
\end{equation}

Using solar XUV luminosity from the input solar spectrum, we calculate $\dot{M}\approx9\times10^9$ $\rm g\;s^{-1}$. We can also take the XUV luminosity for HAT-P-32 reported in \cite{Czesla2022} and calculate $\dot{M}\approx2\times10^{12}$ $\rm g\;s^{-1}$. (Recall that HAT-P-32 is the system very similar to the WASP-48 system.) We therefore expect the mass-loss rate to be on the order of magnitude between $\rm 10^9-10^{12}\;g\;s^{-1}$. We know HAT-P-32 is likely younger than WASP-48, and we therefore can expect the XUV luminosity of WASP-48 to be slightly lower than that of HAT-P-32. Because of this, helium absorption for WASP-48b has been modeled over $10^9-10^{11.5}$ $\rm g$ $\rm s^{-1}$ for mass-loss rate. We also model the thermosphere temperature between 6000--10000 K, which is a typical prediction for close-in, highly irradiated planets.

Helium absorption predicted by the isothermal model at a thermosphere temperature of 9500 K across a range of mass-loss rates is shown in Figure \ref{model_compare_ex}. While keeping all stellar parameters constant, as mass-loss rate increases, so does the absorption depth of the helium feature. We can compare this to Figure 4 in \citealt{Lampon2020}, which demonstrates how the density of the metastable state increases with increasing mass-loss rate. This may occur because an increasing mass-loss rate increases corresponds to an increase in atmospheric density and thus free electron density (\citealt{Oklopcic2019}; also see Figure 6 in \citealt{Lampon2020}). Because free electrons are responsible for recombining ionized helium to the metastable triplet state, the helium absorption signature is strengthened. 

Likewise, we can examine how thermosphere temperature drives the absorption feature by plotting a range of temperatures across a single mass-loss rate ($\rm \dot{M}=10^{10}$ $\rm g$ $\rm s^{-1}$). This is shown in Figure \ref{model_compare_ex_temps}. From the figure, there is an inverse relationship between helium absorption and thermosphere temperature. We again compare this with Figure 4 in \citealt{Lampon2020}, which shows a decrease in density of the metastable state with increasing thermosphere temperature. This may be due to the fact that lower temperatures lead to higher atmospheric and electron density by slowing the rate of outflow \citep{Spake2022}. Higher levels of helium ionization due to higher temperatures may also lead to a lower metastable helium abundance.

\begin{figure}
    \hspace*{-0.6cm}
    \centering
    \includegraphics[scale=0.58]{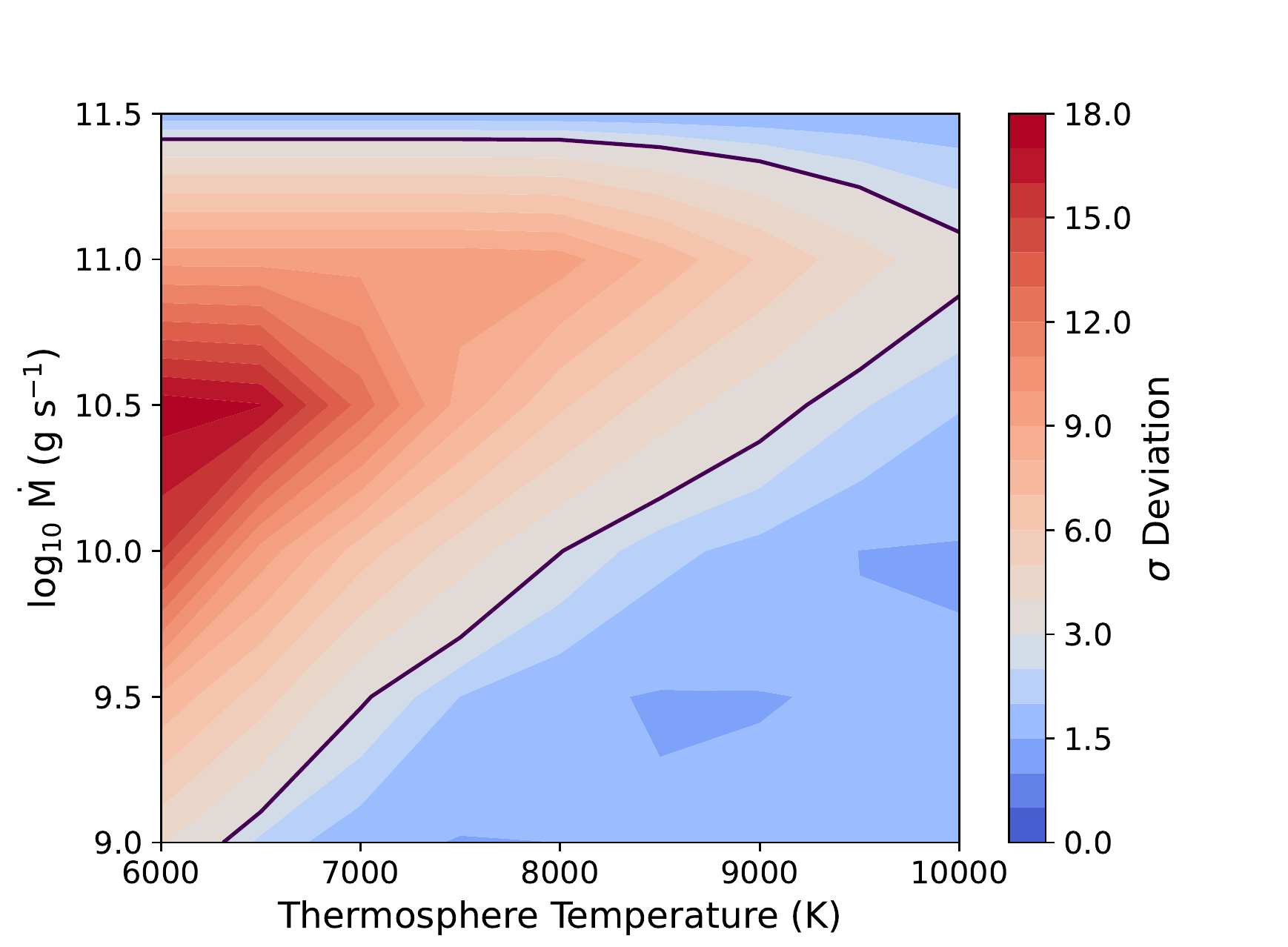}
    \caption{The number of standard deviations between the transmission spectrum and model in the range 10831--10835 \r{A}, measured across the full range of parameter space. A line is drawn around the $3\sigma$ contour to show that anything higher than this value (red on the plot) can be rejected. The apparent decrease in absorption at high mass-loss rates may be a model artifact due to the fact that our model does not take into account helium that exists beyond the Hill radius.}
    \label{models_sig}
\end{figure}

To more fully constrain the parameters of WASP-48b's atmosphere, we create a contour plot depicting the predicted helium absorption signature across the full range of thermosphere temperature and mass-loss rate space. We calculate how well different combinations of thermosphere temperatures and mass-loss rates agree with our observations. This is seen in Figure \ref{models_sig}, which measures the number of $\sigma$ deviations between the transmission spectrum and model across a range of parameter space. The $\chi^2_{\nu}$ value between the data and model is calculated for the range 10831--10835~\r{A}, and this is converted into a $\sigma$ value. Blue contours represent a sigma deviation of less than $3\sigma$, while contours in red represent a sigma deviation of greater than $3\sigma$. In other words, a larger signal is expected in the red regions. Since we do not see evidence of a signal in our data, the areas in red can be rejected.

We can safely reject low thermosphere temperatures ($\sim$ 6000 K) between $\rm \dot{M}\approx10^{9}-10^{11.4}\;g\;s^{-1}$. However, increasing the thermosphere temperature decreases the range of mass-loss rates that can be rejected. As discussed above, low mass-loss rates and high thermosphere temperatures both decrease the atmospheric density, thereby decreasing the metastable helium population. At low thermosphere temperatures but moderate mass loss rates ($\rm \dot{M}=10^{10}-10^{11}\;g\;s^{-1}$), the metastable helium population is maximized \citep{Oklopcic2019}; we see that a smaller signal is expected as thermosphere temperature increases in this mass loss range. At high mass loss rates, the atmospheric density can become too high for the EUV radiation to ionize helium atoms low in the thermosphere, meaning metastable helium only forms in a detached shell that gets thinner and thinner as it approaches the Hill radius. At the highest mass-loss rates, helium mostly forms outside the Hill radius. This is beyond the range considered by our model, which relies on spherical symmetry and therefore breaks down outside the Hill radius. The apparent decrease in absorption at high mass-loss rates therefore may not be entirely physical.




From Figure \ref{models_sig}, we conclude that our non-detection is most likely in agreement with a low mass-loss rate and high thermosphere temperature. In reality, these parameters are correlated with the incident stellar flux \citep{Oklopcic2019}. Further observational and theoretical work is needed to tease out the interplay between them.

\section{Discussion} \label{discussion}

Our non-detection of planetary helium in WASP-48b can be explained through low levels of EUV radiation and atmospheric escape, low helium abundances, or stellar winds and other sources of stellar variability. The most straightforward explanation is the first, but other explanations should be considered. There could still be ongoing atmospheric escape, but the thermosphere temperature may be high enough to ionize a majority of the helium population \citep{Owen2019}. Another possibility is that there is stellar variability due to XUV luminosity variability, stellar winds, shear instability, or stellar flares \citep{Zhang2022}. This variability would lead to varying planetary outflows and thus a time-dependent absorption feature that is only detectable during certain epochs. 

However, though stellar winds could induce a variable helium feature, they are unlikely to suppress it entirely \citep{Zhang2021}. Additionally, while it is possible that high thermosphere temperatures are attenuating the population of metastable helium, the most plausible explanation is that the amount of EUV irradiation by the star is insufficient to support ongoing atmospheric escape. This is supported by the fact that the stellar helium feature seen in Figure \ref{sky_sub} appears weaker than in HAT-P-32 (see Figures C.1 and C.2 in \citealt{Czesla2022}), suggesting that stellar radiation is the main culprit. (Recall that HAT-P-32b, which is very similar to WASP-48b, has a strong helium detection.) Because the same photoionization-recombination pathway that populates the $\rm 2^3S$ metastable state in planetary atmospheres also populates the metastable state in stellar chromospheres, the fact that the stellar signal is weaker in WASP-48 may be an indicator that WASP-48's stellar radiation field is less optimal for the excitation of helium atoms to the metastable state compared to HAT-P-32.

Because EUV radiation decreases with stellar age (e.g., \citealt{Ribas2005, Sanz2011}), this conclusion also makes sense in light of the fact that WASP-48 is an old, slightly evolved star. Indeed, the only other helium paper examining a planet orbiting an evolved star (WASP-12b) was also a non-detection \citep{Kriedberg2018}. Though WASP-48 is rotating rapidly, which would hint at higher stellar activity, we do not see any indication that activity is driving atmospheric escape in WASP-48b. 

\subsection{Helium and EUV Flux}

EUV flux is thought to be the main driver behind metastable helium absorption, but EUV stellar radiation is difficult to observe due to its high absorption by the interstellar medium \citep{Sanz2011}. To date, only one space mission, the \textit{Extreme Ultraviolet Explorer (EUVE)}, has been commissioned for the purpose of studying stellar EUV emission. This leaves astronomers to predict stellar EUV flux by deriving scaling relationships between EUV flux and observable parameters such as age \citep{Sanz2011}, the stellar activity index log $\rm R'_{HK}$ (\citealt{Sreejith2020}), and stellar rotation rate \citep{Wright2011}. 



\cite{Oklopcic2019} also found that it is not simply EUV flux that should correlate with helium absorption, but the ratio of EUV flux to mid-UV flux. While EUV flux populates the $\rm 2^3S$ state by ionizing neutral helium, thereby initializing the photoionization-recombination pathway, the mid-UV flux directly ionizes the $\rm 2^3S$ state and depopulates it. Thus, stars with low mid-UV flux (due to lower effective temperatures) but high EUV flux (due to stellar activity) should optimize helium absorption. \cite{Oklopcic2019} found that planets orbiting K-type stars are the most promising candidates given these constraints.


Recall that WASP-48b orbits an old ($\sim7.9$ Gyr), slightly evolved F star. According to the discussion above, neither F-type stars nor old stars are ideal for the detection of helium. WASP-48's low log $\rm R'_{HK}$ value ($-5.135$) further supports the notion that this is an inactive star, and thus has low amounts of high-energy radiation being emitted from the corona. However, its high rotation rate ($v{\rm\;sin}\;i\sim12.2\; {\rm\;km}\;{\rm\;s^{-1}}$), particularly for a star of its age, suggests that perhaps its activity level could be higher than anticipated. Our non-detection in spite of this high rotation rate hints that perhaps rotation rate is not the best proxy for stellar activity or that age is a more dominant factor.  

Though the non-detection of helium in WASP-48b seems to not raise any questions in light of current understandings of EUV flux, stellar activity, and age, to more robustly investigate these relationships, the full suite of helium detections and non-detections should be considered. Only then can we draw conclusions about the observational veracity of these theoretical assumptions.

\subsection{Comparison with Helium Literature} \label{lit}


In order to tease out the influence of various stellar and planetary parameters, we examine our non-detection in the context of all helium detections and non-detections to date. To the best of our knowledge, WASP-48b is the $ \rm37^{th}$ planet to be investigated for the detection of helium. Table \ref{lit_overview} gives an overview of all the helium detections and non-detections by spectral type. (We utilized both the NASA Astrophysics Data System\footnote{\url{https://ui.adsabs.harvard.edu}} and the IAC ExoAtmospheres database\footnote{\url{http://research.iac.es/proyecto/exoatmospheres/}} to conduct our search.) One can see that there have been more non-detections (21) than detections (16), and for spectral types F, G, and M, the majority of investigations have been non-detections. Only one planet orbiting a hot A-type star has been investigated (KELT-9b), and this was a non-detection \citep{Nortmann2018}. Notably, 18 out of the 37 planets investigated orbit K stars, and of these, more than half have been detections. While this supports the finding from \cite{Oklopcic2019} that K stars are the most promising candidates, the sample is still relatively small and skewed toward K stars to draw any final conclusion. In addition, there have been several non-detections around K stars, as well as detections around other spectral types.
\begin{deluxetable}{ccc}
\tablecaption{Distribution of helium detections and non-detections in the literature by spectral type. \label{lit_overview}}
\tablenum{4}
\tablewidth{0pt}
\def\arraystretch{.85}
\tablehead{\colhead{Spectral Type} & \colhead{Detections} & \colhead{Non-Detections}}
\startdata
         A & 0 & 1\\
         F & 1 & 3\\
         G & 1 & 4\\
         K & 12 & 6\\
         M & 1 & 6\\
         T Tauri & 1 & 1 \\
         \textbf{Total} & \textbf{16} & \textbf{21} \\
\enddata
\end{deluxetable}

Still, the majority of the findings around F, G, and M stars have thus far been non-detections. The sole detection around an F star was for the hot Jupiter HAT-P-32b \citep{Czesla2022}, which, as noted, is very similar to the WASP-48 system, albeit likely younger. 

The full table of stellar and planetary parameters for each planet are found in Table \ref{stellar_params}. These parameters have been taken from the literature. To the best of our knowledge, this is the most complete sample of helium literature to date. While there are inhomogeneities in the way that the data are derived (e.g., some stellar ages are derived through isochrone fitting, while others are derived using gyrochronology), this can be used as a starting point to compare various parameters with helium absorption. 

To compare helium findings across the literature, we measure the height of the planetary atmosphere following \cite{Fossati2022}, in which the effective planet radius ($R_{\rm{eff}}$) at 10833 \r{A} (in units of $R_P$) is calculated from
\begin{equation}
    \frac{R_{\rm{eff}}}{R_P}=\sqrt{\frac{\delta+c}{\delta}},
\end{equation}

\noindent where $R_P$ is the planet radius, $\delta$ is the transit depth, and $c$ is the helium absorption depth or upper limit. From here, we calculate a change in transit depth, $\delta_{R_P}$, which represents the height of an opaque atmosphere that causes the inflated planetary effective radius at 10833 \r{A} and produces the absorption signature in the transmission spectrum \citep{Nortmann2018}. $\delta_{R_P}$ is given by
\begin{equation}
    \delta_{R_P}=\frac{R_{\rm{eff}}}{R_P}-1.  
\end{equation}

\noindent $\delta_{R_P}$ is then normalized to the scale height, $H_{eq}$, following \cite{Nortmann2018}. To calculate $H_{eq}$, we use the planetary equilibrium temperature, $T_{eq}$, and set the atmospheric mean molecular weight to 2.3 times the mass of hydrogen, assuming H/He planetary atmospheres with solar abundances.

We examine the correlation between the normalized atmospheric height of the helium signature ($\delta_{R_P}/H_{eq}$) and several stellar and planetary parameters, including the stellar activity index log $\rm R'_{HK}$, stellar age, stellar rotation rate, metallicity, planetary surface gravity, and semimajor axis. We exclude directly comparing $\delta_{R_P}/H_{eq}$ with high-energy EUV or XUV flux, as has been done in several other helium papers (e.g., \citealt{Nortmann2018, Alonso2019, Kasper2020, Casa2021, Orell-Miquel2022, Fossati2022}), because the X-ray luminosity for WASP-48 is not available. Instead of moving through multiple layers of analytic scaling relations to estimate the X-ray luminosity and in turn the EUV flux onto the planet, we elect instead to compare $\delta_{R_P}/H_{eq}$ with only directly measured parameters found in the literature. 

We do not see any correlation with the stellar and planetary parameters listed above, with the exception of the stellar activity index log $\rm R'_{HK}$. log $\rm R'_{HK}$ was first described in \cite{Noyes1984}, and measures the ratio of the emission of Ca II H and K lines (3933 \r{A} and 3968 \r{A}) originating in the chromosphere compared to the total stellar bolometric luminosity. The correlation between log $\rm R'_{HK}$ and $\delta_{R_P}/H_{eq}$ was first noted in \cite{Nortmann2018}, and we find that using log $\rm R'_{HK}$ is advantageous because it is easily found in the literature and involves optical emission lines that can be measured from the ground \citep{Sreejith2020}. However, because the $\rm Ca~II$ lines originate in the chromosphere, they do probe a spatially independent region from the corona and transition region, where EUV radiation originates \citep{Nortmann2018, Sreejith2020}. Additionally, this measurement is not available for all stellar types, as late-type stars have very low luminosity in the optical region. Still, log $\rm R'_{HK}$ is widely used as a method to quantify the stellar magnetic activity. 

\begin{figure*}
    \centering
    \hspace*{-0.8cm}
    {\includegraphics[scale=0.85]{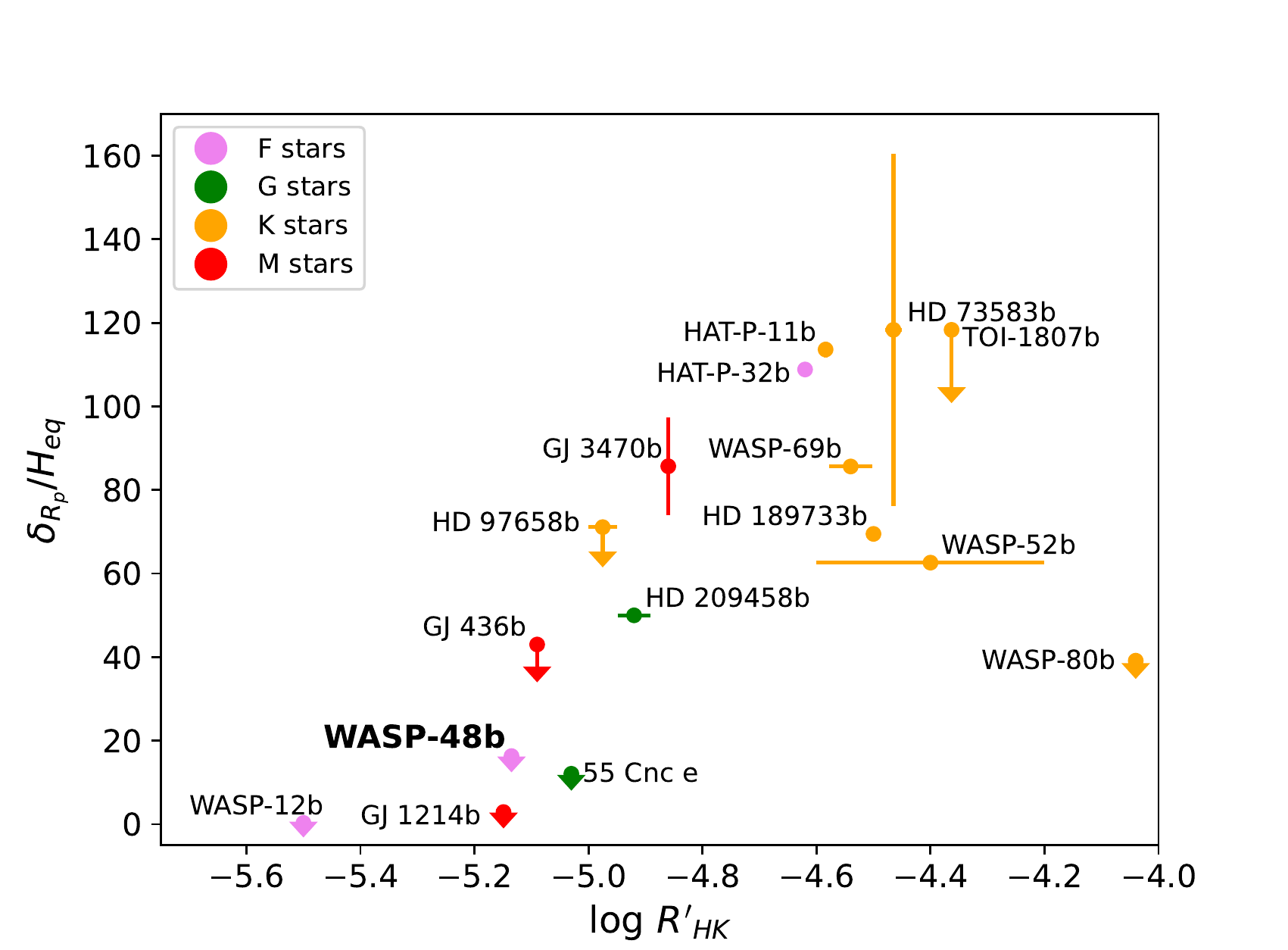}}
    \caption{Normalized helium absorption versus log $\rm R'_{HK}$, with spectral type designated by color. Vertical error bars are depicted for all data points, though some are quite small and are not visible. Error bars on log $\rm R'_{HK}$ are included where available. Non-detections (upper limits on helium absorption) are depicted with vertical arrows. There appears to be a slight positive correlation between $\delta_{R_P}/H_{eq}$ and log $\rm R'_{HK}$, hinting at a relationship between helium absorption and stellar activity. WASP-48b is in the lower left corner at a $\delta_{R_P}/H_{eq}$ value of $\sim16.3$ and log $\rm R'_{HK}$ value of $-5.135$.}
    \label{logRHK}
\end{figure*}

We plot $\delta_{R_P}/H_{eq}$ vs. log $\rm R'_{HK}$ for those stars in our sample with available log $\rm R'_{HK}$ values in the literature in Figure \ref{logRHK}. We include only planets for which helium absorption is derived spectroscopically, as photometric measurements typically have a diluted $\delta_{R_P}$ due to the narrowband filter's response function. They therefore require a scaling factor in order to be accurately compared with spectroscopic measurements. 

In Figure \ref{logRHK}, spectral types are delineated by color, and non-detections are marked by a vertical downward arrow. At first glance, there does appear to be some positive correlation between these two parameters. Using a Pearson correlation coefficient to measure the linear correlation between these parameters for detections only, we find a correlation coefficient of $+0.29$, indicating a weak to moderate positive correlation. 

Still, Figure \ref{logRHK} must be interpreted with a grain of skepticism. Because we calculate $\delta_{R_P}/H_{eq}$ using values taken from various literature sources, this is in no way a homogeneous sample. Helium absorption depths are measured differently using different telescopes, and several assumptions go into the predicted scale height of an atmosphere. Additionally, applying a scale height normalization may actually obscure the relationship between $\delta_{R_P}/H_{eq}$ and $\rm R'_{HK}$ if the scale height, which is related to the lower hydrostatic atmosphere, is uncorrelated to the height of the thermosphere. Both detections and non-detections have been reported for the same planet (e.g., helium in WASP-52b was not detected in \citealt{Vissapragada2020} but was detected in \citealt{Kirk2022}), and absorption values are constantly updated. We must keep these factors in the back of our mind when interpreting such a plot. A greater sample size with more diversity in spectral type will help us uncover whether there is truly a relationship between log $\rm R'_{HK}$ values and helium absorption. 

Indeed, while many papers report a positive trend between helium absorption and EUV or XUV flux (e.g., \citealt{Nortmann2018, Poppenhaeger2022}), others are beginning to question whether this simple correlation is the only factor or whether the relationship is shallower than previously assumed (e.g., \citealt{Alonso2019, Fossati2022, Kirk2022}). Next steps should consider comparing the "hardness" of the high-energy stellar spectra, i.e. the ratio between the EUV and mid-UV flux as described in \cite{Oklopcic2019}. 

\section{Conclusion} \label{conclusion}

In this paper, we search for metastable helium absorption in the extended atmosphere of WASP-48b, a hot Jupiter orbiting a slightly evolved, rapidly rotating F-type star. We use high-resolution ground-based transmission spectroscopy using the Habitable-Zone Planet Finder. Our resulting transmission spectrum is flat. No planetary helium appears to be present in the extended atmosphere of WASP-48b, suggesting that there is no ongoing atmospheric escape occurring on this planet.

We compare our findings with a 1D isothermal Parker wind model \citep{Oklopcic2019}, which allows us to constrain the expected thermosphere temperature and mass-loss rate of the planet. We find that our non-detection can be explain by a high thermosphere temperature and low mass-loss rate, or a high mass-loss rate.

To understand this non-detection to the fullest extent, we compare our findings to all other published helium works. We normalize helium absorption to the planetary atmospheric scale height, $H_{eq}$, and plot these values against a variety of stellar and planetary parameters. We find that only the log $\rm R'_{HK}$ index shows any correlation with helium absorption. 

\cite{Oklopcic2019} predicts that close-in planets orbiting K stars are the most likely to experience high levels of atmospheric escape and provide the most promising candidates for helium absorption. While there has certainly been the largest number of detections around K stars, there have also been detections around other spectral types (F, G, M, and T Tauri stars) as well as several non-detections around K stars. More investigations of planets around a variety of stellar types and ages need to be done to observationally validate the prediction that K-type stars are optimal for helium detections.

To strengthen the veracity of atmospheric escape studies, we should couple helium observations with Ly-$\alpha$ (for stars closer than ~50 pc) and/or H$\alpha$ studies to corroborate findings when possible. This has been done for some planets, including HD 189773b \citep{Vidal-Madjar2003, Jensen2012, Salz2018}, 55 Cnc e \citep{Ehrenreich2012, Zhang2021}, and GJ 9827b and d \citep{Carleo2021}. Additionally, multiple transits should be observed when possible, because it has become increasingly apparent that stellar variability plays a meaningful role in fluctuating mass loss and helium absorption over short timescales. HD 189733b has been shown to exhibit XUV variability of up to 33\% \citep{Pillitteri2022, Zhang2022}, which greatly impacts the amount of helium absorption detected. 

As more planets are investigated and comparative studies expand, one of the fundamental goals of the exoplanet atmospheric research community must be to fully elucidate the timeline of atmospheric escape, the stellar and planetary environment in which it occurs, and its impact on atmospheric evolution over a planet's lifetime. This is just one small piece of the exoplanet characterization puzzle, but it will bring us closer to understanding the architecture and habitability of planetary systems within our Galaxy.


\movetabledown=2.75in
\begin{rotatetable*}
\begin{deluxetable*}{cccccccccc}
\tablecaption{Stellar parameters for exoplanets with published helium results. \label{stellar_params}}
\tablenum{5}
\tablewidth{0pt}
\def\arraystretch{.85}
\tablehead{\colhead{Planet} & \colhead{Distance (pc)} & \colhead{Spectral Type} & \colhead{$T_{\rm eff}$ (K)} & \colhead{Age (Gyr)} & \colhead{Radius ($R_{\odot}$)} & \colhead{Mass ($ M_{\odot}$)} & \colhead{[Fe/H]} & \colhead{$v$ $\rm sin$ $i$ ($\rm km$ $\rm s^{-1}$)} & \colhead{$\rm log$ $\rm R^{'}_{HK}$}}
\startdata
        WASP-107b & $64.7414{{^{+0.2617}_{-0.2596}}}$ & $\rm K6V^1$ & $4425\pm70^2$ & $6.9^{+3.7 ^ 2}_{-3.4}$ & $0.67\pm0.02^2$ & $0.683^{+0.017 ^ 2}_{-0.016}$ & $+0.02\pm0.09^2$ & $2.5\pm0.8^1$ & $-$ \\
        WASP-69b & $49.9605^{+0.1320}_{-0.1313}$ & $\rm K5V^4$ & $4700\pm50^4$ & $1.10\pm0.15^{4}$ & $0.813\pm0.028^4$ & $0.826\pm0.029^4$ & $+0.15\pm0.08^4$ & $2.2\pm0.4^4$ & $-4.54^4$\\
        HAT-P-11b & $37.7647^{+0.0337}_{-0.0336}$ & $\rm K4V^6$ & $4850\pm50^6$ & $6.5^{+5.9 ^ 6}_{-4.1}$ & $0.75\pm0.02^6$ &  $0.81^{+0.02 ^ 6}_{-0.03}$ & $+0.31\pm0.05^6$ & $1.5\pm1.5^6$ & $−$ $-4.584^6$\\
        HD 189733b & $19.7638{^{+0.0128}_{-0.0127}}$ & $\rm K2V^8$ & $5050\pm50^8$ & $5.2\pm3.5^9$ & $0.76\pm0.01^8$ & $0.82\pm0.03^8$ & $-0.03\pm0.04^8$ & $3.5\pm1.0^8$ & $-4.5\pm0.007^{75}$\\
        WASP-12b & $427.246^{+6.068}_{-5.903}$ & $\rm F^{13}$ & $6300\pm{200}$ & $2\pm1^{13}$ & $1.57\pm0.07^{13}$ & $1.35\pm0.14^{13}$ & $+0.30^{14}$ & $2.2\pm1.5^{13}$ & $-5.500^{14}$\\
        HAT-P-18b & $161.400^{+0.610}_{-0.605}$ & $\rm K2V^{16}$ & $4803\pm80^{16}$ & $12.4^{+4.4 ^ {16}}_{-6.4}$ & $0.749\pm0.037^{16}$ & $0.770\pm0.031^{16}$ & $+0.10\pm0.08^{16}$ & $0.5\pm0.5^{16}$ & $-4.73^{77}$\\
        HD 209458b & $48.3016^{+0.1240}_{-0.1234}$ & $\rm G0V^{18}$ & $6071\pm20^{18}$ & $3.5\pm1.4^{18}$ & $1.20\pm0.05^{18}$ & $1.26\pm0.15^{18}$ & $+0.02\pm0.03^{20}$ & $3.75\pm1.25^{21}$ & $–4.92\pm0.029^{75}$\\
        55 Cnc e & $12.5855^{+0.0124}_{-0.0123}$ & $\rm G8V^{25}$ & $5172\pm18^{26}$ & $8.6\pm1.0^{27}$ & $0.943\pm 0.010^{27}$ & $0.905\pm0.015^{27}$ & $+0.35\pm0.1$ & $<1.23\pm0.01^{27}$ & $–5.03^{27}$ \\
        GJ 1214b & $14.6427\pm0.0372$ & $\rm M4.5V^{29}$ & $3250\pm100^{29}$ & $5-10^{30}$ & $0.2213\pm0.0043^{29}$ &  $0.176\pm0.009^{29}$ & $+0.1\pm0.1^{29}$ & $<2.0^{31}$ & $-5.149^{85}$\\
        GJ 9827d & $29.6610^{+0.0543}_{-0.0541}$ & $\rm K6V^{34}$ & $4255\pm110^{34}$ & $~10^{49}$ & $0.651\pm0.065^{34}$ & $0.659\pm0.060^{34}$ & $-0.28\pm0.12^{34}$ & $2\pm1^{34}$ & $-$\\
        HD 97658b & $21.5618^{+0.0254}_{-0.0253}$ & $\rm K1V^{36}$ & $5192\pm122^{36}$ & $3.9\pm2.6^{37}$ & $0.746\pm0.034^{36}$ & $0.77\pm0.05^{36}$ & $-0.23\pm0.03^{38}$ & $0.5\pm0.5^{38}$ & $-4.975\pm0.025^{38}$ \\
        GJ 436b & $9.75321^{+0.00898}_{-0.00896}$ & $\rm M2.5V^{39}$ & $3479\pm60^{40}$ & $6.5^{39}$ & $0.449\pm0.019^{40}$ & $0.445\pm0.044^{40}$ & $-0.03\pm0.20^{23}$ & $0.33\pm0.091^{40}$ & $-5.09\pm0.001^{75}$\\
        V1298b & $108.199^{+0.704}_{-0.696}$ & T $\rm Tauri^{42}$ & $4970\pm120^{42}$ & $0.023\pm0.004^{42}$ & $1.345^{+0.056 ^ {42}}_{-0.051}$ & $1.101^{+0.049 ^ {42}}_{-0.051}$ & $-$ & $-$ & $-$\\
        V1298d & $108.199^{+0.704}_{-0.696}$ & T $\rm Tauri^{42}$ & $4970\pm120^{42}$ & $0.023\pm0.004^{42}$ & $1.345^{+0.056 ^ {42}}_{-0.051}$ & $1.101^{+0.049 ^ {42}}_{-0.051}$ & $-$ & $-$ & $-$\\
        GJ 3470b & $29.4214^{+0.0508}_{-0.0507}$ & $\rm M1.5V^{44}$ & $3652\pm50^{44}$ & $0.3-3^{44}$ & $0.48\pm0.04^{44}$ & $0.51\pm0.06^{44}$ & $+0.20\pm0.10^{44}$ & $-$ & $-4.86\pm0.0038^{75}$\\
        KELT-9b & $204.455^{+1.582}_{-1.558}$ & $\rm A0V^{47}$ & $10170\pm450^{47}$ & $0.3^{47}$ & ${2.362^{+0.075}_{-0.063}}^{ 47}$ & ${2.52^{+0.25}_{-0.20}}^{47}$ & $-0.030\pm0.200^{47}$ & $111.4\pm1.3^{47}$ & $-$\\
        GJ 9827b & $29.6610^{+0.0543}_{-0.0541}$ &$\rm K6V^{34}$ & $4255\pm110^{34}$ & $10^{49}$ & $0.651\pm0.065^{34}$ & $0.659\pm0.060^{34}$ & $-0.28\pm0.12^{34}$ & $2\pm1^{34}$ & $-$\\
        WASP-80b & $49.78760\pm0.11605$ & $\rm K7V^{51}$ & $4150\pm100^{10}$ & $1.601\pm0.202^{52}$ & $0.571\pm0.016^{10}$ & $0.570\pm0.050^{10}$ & $-0.14\pm0.16^{51}$ & $3.46\pm0.35^{51}$ & $-4.04^{54}$ \\
        WASP-76b & $194.459^{+6.206}_{-5.839}$ & $\rm F7V^{55}$ & $6250\pm100^{55}$ & $5.3\pm2.9^{55}$ & $1.73\pm0.04^{55}$ & $1.46\pm0.07^{55}$ & $+0.23\pm0.10^{55}$ & $3.3\pm0.6^{55}$ & $-$\\
        TRAPPIST-1b & $12.47\pm0.01$ & $\rm M8V^{57}$ & $2566\pm26^{58}$ & $7.6\pm2.2^{59}$ & $0.1192\pm0.0013^{58}$ & $0.0898\pm0.0023^{58}$ & $+0.040\pm0.080^{57}$ & $6\pm2^{57}$ & $-$ \\
        TRAPPIST-1e & $12.47\pm0.01$ & $\rm M8V^{57}$ & $2566\pm26^{58}$ & $7.6\pm2.2^{59}$ & $0.1192\pm0.0013^{58}$ & $0.0898\pm0.0023^{58}$ & $+0.040\pm0.080^{57}$ & $6\pm2^{57}$ & $-$\\
        TRAPPIST-1f & $12.47\pm0.01$ & $\rm M8V^{57}$ & $2566\pm26^{58}$ & $7.6\pm2.2^{59}$ & $0.1192\pm0.0013^{58}$ & $0.0898\pm0.0023^{58}$ & $+0.040\pm0.080^{57}$ & $6\pm2^{57}$ & $-$\\
        HD 63433b & $22.4035\pm0.0225$ & $\rm G2V^{62}$ & $5640\pm74^{62}$ & $0.414\pm0.023^{62}$ & $0.912\pm0.034^{62}$ & $0.990\pm0.030^{62}$ & $+0.04\pm0.05^{62}$ & $7.3\pm0.3^{62}$ & $-4.39\pm0.05^{62}$ \\
        HD 63433c & $22.4035\pm0.0225$ & $\rm G2V^{62}$ & $5640\pm74^{62}$ & $0.414\pm0.023^{62}$ & $0.912\pm0.034^{62}$ & $0.990\pm0.030^{62}$ & $+0.04\pm0.05^{62}$ & $7.3\pm0.3^{62}$ & $-4.39\pm0.05^{62}$ \\
        HAT-P-32b & $289.205^{+5.355}_{-5.167}$ & $\rm F^{64}$ & $6207\pm88^{10}$ & $3.8\pm1.5^{64}$ & $1.219\pm0.016^{64}$ & $1.160\pm0.041^{64}$ & $-0.04\pm0.08^{64}$ & $20.7\pm0.5^{64}$ & $-4.62^{64}$ \\
        HD 73583b & $31.5666^{+0.0321}_{-0.0320}$ & $\rm K4V^{66}$ & $4511\pm110^{66}$ & $0.48\pm0.19^{66}$ & $0.65\pm0.02^{66}$ & $0.73\pm0.02^{66}$ & $0.00\pm0.09^{66}$ & $3.5\pm0.5^{66}$ & $-4.465\pm0.015^{66}$ \\
        TOI-2136b & $33.36310\pm0.06425$ & $\rm M4.5V$ & $3373\pm108^{68}$ & $4.6\pm±1.0^{69}$ & $0.3440\pm0.0099^{68}$ & $0.3272\pm0.0082^{68}$ & $+0.02\pm0.14^{68}$ & $0.21\pm0.012^{68}$ & $-$\\
        WASP-52b & $174.818^{+1.343}_{-1.323}$ & $\rm K2V^{70}$ & $5000\pm100^{70}$ & $0.4\pm0.3^{70}$ & $0.79\pm0.02^{70}$ & $0.87\pm0.03^{70}$ & $+0.03\pm0.12^{70}$ & $3.6\pm0.9^{70}$ & $-4.4\pm0.2^{70}$ \\
        WASP-127b & $159.507^{+1.210}_{-1.191}$ & $\rm G5V^{72}$ & $5620\pm85^{72}$ & $11.41\pm1.80^{72}$ & $1.39\pm0.03^{72}$ & $1.08\pm0.03^{72}$ & $-0.18\pm0.06^{72}$ & $0.3\pm0.2^{72}$ & $-$\\
        WASP-177b & $176.818^{+2.428}_{-2.364}$ & $\rm K2V^{74}$ & $5017\pm70^{74}$ & $2.5\pm1.7^{74}$ & $0.885\pm0.046^{74}$ & $0.876\pm0.038^{74}$ & $0.25\pm0.04^{74}$ & $2.9\pm0.2^{74}$ & $-$ \\
        HAT-P-26b & $141.837^{+1.152}_{-1.133}$ & $\rm K1V^{76}$ & $5079\pm88^{76}$ & $9.0^{+3.0 ^ {76}}_{-4.9}$ & $0.788^{+0.098 ^ {76}}_{-0.043}$ & $0.816\pm0.033^{76}$ & $-0.04\pm0.08^{76}$ & $1.8\pm0.5^{76}$ & $-4.992^{76}$\\
        NGTS-5b & $306.779^{+2.601}_{-2.558}$ & $\rm K2V^{78}$ & $4987\pm41^{78}$ & $-$ & $0.739^{+0.014\;78}_{-0.012}$ & $0.661^{+0.068\;78}_{-0.061}$ & $0.12\pm0.10^{78}$ & $-$ & $-4.63^{77}$\\
        TOI-1807b & $42.5775^{+0.0622}_{-0.0620}$ & $\rm K3V^{79}$ & $4730\pm75^{80}$ & $0.3\pm0.08^{80}$ & $0.690\pm0.036^{80}$ & $0.76\pm0.03^{80}$ & $-0.04\pm0.02^{80}$ & $4.2\pm0.5^{80}$ & $-4.363\pm0.002^{80}$ \\
        TOI-2076b & $41.9091^{+0.0528}_{-0.0526}$ & $\rm K0V^{79}$ & $5187^{+54\;81}_{-53}$ & $0.2\pm0.05^{81}$ & $0.7622^{+0.0157\;81}_{-0.0159}$ & $0.850^{+0.025\;81}_{-0.026}$ & $-0.032^{+0.048\;81}_{-0.047}$ & $4.3\pm0.5^{81}$ & $-4.271\pm0.056^{81}$ \\
        TOI-1430.01 & $41.17\pm0.04$ & $\rm K1V^{82}$ & $5067\pm60^{82}$ & $0.165\pm0.03^{82}$ & $0.784^{+0.018 ^ {82}}_{-0.014}$ & $0.85\pm0.10^{82}$ & $-$ & $6.9^{82}$ & $-$ \\
        TOI-1683.01 & $51.19\pm0.14^{82}$ & $\rm K4V^{82}$ & $4539\pm100^{82}$ & $0.5\pm0.15^{82}$ & $0.636^{+0.031 ^ {82}}_{-0.017}$ & $0.69\pm0.09^{82}$ & $-$ & $2.8^{82}$ & $-$ \\
        WASP-48b & $454.144^{+4.465}_{-4.381}$ & $\rm F^{83}$ & $5920\pm150^{83}$ & $7.9^{+2.0\;83}_{-1.6}$ & $1.75\pm0.09^{83}$ & $1.19\pm0.05^{83}$ & $-0.12\pm0.12^{83}$ & $12.2\pm0.7^{83}$ & $-5.135^{84}$\\
\enddata
\tablecomments{\scriptsize{Values taken from NASA Exoplanet Archive\footnote{\url{https://exoplanetarchive.ipac.caltech.edu}} unless otherwise noted. Error bars are taken from the literature or, if missing, estimated by taking the average error from the remaining data. See page 16 for references.}} 
\end{deluxetable*}
\end{rotatetable*}

\movetabledown=2.25in
\begin{rotatetable*}
\begin{deluxetable*}{ccccccccc}
\tablecaption{Planetary parameters for exoplanets with published helium results. \label{plan_params}}
\tablenum{5}
\tablewidth{0pt}
\def\arraystretch{.85}
\tablehead{\colhead{Planet} & \colhead{Radius ($R_J$)} & \colhead{Mass ($M_J$)} & \colhead{$a$ (au)} & \colhead{$T_{\rm eq}$ (K)} & \colhead{$\rm log$ $g_P$ ($\rm cm\;s^{-2}$)} & \colhead{$\delta_{R_P}$} & \colhead{$\delta_{R_P}/H_{eq}$} & \colhead{Detection?}}
\startdata
        WASP-107b & $0.94\pm0.02^1$ & $0.096\pm0.005^2$ & $0.055\pm0.001^1$ & $770\pm60^1$ & $2.49\pm0.05^1$ & $1.12\pm0.10^{3}$ & $84.2\pm6.6^{3}$ & Yes\\
        WASP-69b & $1.057\pm0.047^4$ & $0.260\pm0.017^4$ & $0.04525\pm0.00053$ & $963\pm18$ & $2.726\pm0.046^4$ & $0.74\pm0.13^5$ & $85.6\pm1.6^5$ & Yes\\
        HAT-P-11b & $0.422\pm0.014^6$ & $0.081\pm0.009^6$ & $0.0530^{+0.0002}_{-0.0008}$ & $878\pm15$ & $3.05\pm0.06^6$ & $1.06\pm0.11^7$ & $113.6\pm1.9^7$ & Yes\\
        HD 189733b & $1.138\pm0.027^{10}$ & $1.123\pm0.045^{10}$ & $0.0313\pm0.0004^8$ & $1201\pm13^{23}$ & $3.3\pm0.03^{10}$ & $0.17\pm0.06^{12}$ & $69.5\pm0.8^{12}$ & Yes\\
        WASP-12b & $1.825\pm0.091^{10}$ & $1.39\pm0.12^{10}$ & $0.02312^{+0.00094 ^ {10}}_{-0.00100}$ & $2516\pm36^{13}$ & $2.99\pm0.03^{13}$ & $0.002\pm0.134^{15}$ & $0.29\pm0.004^{15}$ & No \\
        HAT-P-18b & $0.995\pm0.052^{16}$ & $0.197\pm0.013^{16}$ & $0.0559\pm0.0007^{16}$ & $852\pm28^{16}$ & $2.69\pm0.05^{16}$ & $0.12\pm0.15^{17}\tablenotemark{*}$ & $13.3\pm0.4^{17}\tablenotemark{*}$ & Yes \\
        HD 209458b & $1.359^{+0.016 ^ {10}}_{-0.019}$ & $0.682^{+0.014 ^ {10}}_{-0.015}$ & $0.0490\pm0.0020^{18}$ & $1459\pm12^{22}$ & $2.963\pm0.005^{23}$ & $0.29\pm0.10^{24}$ & $50.0\pm0.4^{24}$ & Yes\\
        55 Cnc e & $0.1673\pm0.00287^{27}$ & $0.0251\pm0.00104^{27}$ & $0.01544\pm0.00005^{27}$ & $1990^{28}$ & $3.35\pm2.00$ & $0.32\pm0.10^{28}$ & $12.1\pm0.6^{28}$ & No\\
        GJ 1214b & $0.254\pm0.018^{32}$ & $0.0197\pm0.0027^{32}$ & $0.01411\pm0.00032^{32}$ & $604\pm19^{29}$ & $2.88\pm2.07$ & $0.046\pm0.148^{33}$ & $2.9\pm0.5^{33}$ & No \\
        GJ 9827d & $0.185\pm0.0125^{35}$ & $0.016^{+0.006}_{-0.004}$ & $0.05615\pm0.00091^{35}$ & $680\pm25^{35}$ &  $3.07\pm0.12^{35}$ & $0.34\pm0.27^{33}$ & $21.4\pm0.8^{33}$ & No \\
        HD 97658b & $0.2055\pm0.009814^{36}$ & $0.0246\pm0.00173^{36}$ & $0.0831\pm0.0011^{38}$ & $751\pm12^{37}$ & $3.16\pm2.09$ & $0.90\pm0.26^{33}$ & $71.1\pm6.1^{33}$ & No\\
        GJ 436b & $0.366\pm0.014^{41}$ & $0.0799^{+0.0066 ^ {40}}_{-0.0063}$ & $0.0308\pm0.0013^{41}$ & $649\pm60^{23}$ & $3.17\pm2.13$ & $0.26\pm0.14^{5}$ & $43.0\pm5.6^{5}$ & No\\
        V1298b & $0.916^{+0.052 ^ {42}}_{-0.047}$ & $-$ & $0.1688\pm0.0026^{42}$ & $677\pm22^{42}$ & $-$ & $0.036\pm0.218^{43}$ & $-$ & No \\
        V1298d & ${0.572^{+0.040}_{-0.035}}^{ 42}$ & $-$ & $0.1083\pm0.0017^{42}$ & $845\pm27^{42}$ & $-$ & $0.10\pm12^{43}$ & $-$ & Yes \\
        GJ 3470b & $0.346\pm0.029^{44}$ & ${0.03958^{+0.00412}_{-0.00403}}^{ 44}$ & $0.0348\pm0.0014^{45}$ & $615\pm16^{45}$ & $2.91\pm2.04$ & $0.93\pm0.32^{46}$ & $85.7\pm11.7^{46}$ & Yes \\
        KELT-9b & ${1.891^{+0.061}_{-0.053}}^{ 47}$ & $2.88\pm0.84^{47}$ & $0.03462\pm0.00110^{47}$ & $4050\pm180^{47}$ & $3.30\pm0.11^{47}$ & $0.22\pm0.11^5$ & $40.8\pm1.8^{5}$ & No \\
        GJ 9827b & ${0.1407^{+0.0024}_{-0.0028}}^{ 48}$ & $0.0154\pm0.0015^{48}$ & ${0.01880^{+0.00020}_{-0.00014}}^{48}$ & $1114\pm46^{49}$ & $3.29\pm2.12$ & $1.66\pm0.82^{50}$ & $81.5\pm6.5^{50}$ & No \\
        WASP-80b & ${0.952^{+0.026}_{-0.027}}^{ 10}$ & ${0.540^{+0.036}_{-0.035}}^{ 10}$ & ${0.03427^{+0.00096}_{-0.00100}}^{10}$ & $825\pm19^{53}$ & $3.178\pm0.013^{51}$ & $0.11\pm0.08^{54}$ & $39.1\pm0.9^{54}$ & No \\
        WASP-76b & $1.83\pm0.06^{53}$ & $0.92\pm0.03^{53}$ & $0.0330\pm0.0005^{53}$ & $2160\pm40^{53}$ & $2.80\pm0.02^{53}$ & $0.32\pm0.10^{56}$ & $34.2\pm0.6^{56}$ & No \\
        TRAPPIST-1b & ${0.09956^{+0.00125}_{-0.00107}}^{ 58}$ & $0.004323\pm0.000217^{58}$ & $0.01154\pm0.00010^{58}$ & $400\pm9^{57}$ & $3.03\pm1.71^{58}$ & $0.040\pm0.115^{60}$ & $2.1\pm0.1^{60}$ & No \\
        TRAPPIST-1e & ${0.0821^{+0.0012}_{-0.0011}}^{ 58}$ & $0.00218\pm0.00007^{58}$ & $0.02925\pm0.00250^{58}$ & $251.3\pm4.9^{61}$ & $2.90\pm1.38^{58}$ & $0.070\pm0.161^{60}$ & $3.6\pm0.1^{60}$ & No \\
        TRAPPIST-1f & ${0.09323^{+0.00116}_{-0.00107}}^{ 58}$ & $0.003269\pm0.000098^{58}$ & $0.03849\pm0.00033^{58}$ & $219.0\pm4.2^{61}$ & $2.97\pm1.38^{58}$ & $0.020\pm0.132^{60}$ & $1.6\pm0.1^{60}$ & No \\
        HD 63433b & $0.192\pm0.009^{62}$ & $-$ & ${0.0719^{+0.0031}_{-0.0044}}^{ 62}$ & $969\pm13^{54}$ & $-$ & $2.42\pm0.57^{63}$ & $-$ & No \\
        HD 63433c & $0.238\pm0.011^{62}$ & $-$ & ${0.1458^{+0.0062}_{-0.0101}}^{ 62}$ & $680\pm9^{54}$ & $-$ & $1.82\pm0.46^{63}$ & $-$ & No \\
        HAT-P-32b & $1.789\pm0.025^{10}$ & $0.75\pm0.13^{10}$ & ${0.03427^{+0.00040}_{-0.00042}}^{ 10}$ & $1786\pm26^{64}$ &  $2.82\pm0.10^{64}$ & $0.83\pm0.05^{65}$ & $108.8\pm1.6^{65}$ & Yes \\
        HD 73583b & $0.249\pm0.009^{66}$ & ${0.0321^{+0.0107}_{-0.0098}}^{ 66}$ & ${0.0604^{+0.0027}_{-0.0026}}^{ 66}$ & $714\pm21^{66}$ & $3.11\pm2.66^{66}$ & $1.32\pm0.17^{67}$ & $118.3\pm42.1^{67}$ & Yes \\ 
        TOI-2136b & $0.20\pm0.0062^{68}$ & $0.015\pm0.0098^{68}$ & $0.0533\pm0.0015^{68}$ & $378\pm13^{68}$ & $2.98\pm2.8^{68}$ & $1.25\pm0.89^{68}$ & $125.3\pm82.9^{68}$ & No \\
        WASP-52b & $1.27\pm0.03^{70}$ & $0.46\pm0.02^{70}$ & $0.0272\pm0.0003^{70}$ & $1315\pm35^{70}$ & $2.81\pm0.03^{70}$ & $0.50\pm{0.08}^{71}$ & $62.6\pm1.7^{71}$ & Yes \\
        WASP-127b & $1.37\pm0.04^{72}$ & $0.18\pm0.02^{72}$ & $0.0520\pm0.0005^{72}$ & $1400\pm24^{72}$ & $2.33\pm0.06^{72}$ & $0.19\pm0.17^{73}$ & $7.8\pm0.1^{73}$ & No\\
        WASP-177b & ${1.58^{+0.66}_{-0.36}}^{74}$ & $0.508\pm0.038^{74}$ & $0.03957\pm0.00058^{74}$ & $1142\pm32^{74}$ & $2.67\pm0.31^{74}$ & $0.17\pm{0.85}^{71}$ & $22.6\pm0.6^{71}$ & No\\
        HAT-P-26b & $0.565^{+0.072\;76}_{-0.032}$ & $0.059\pm0.007^{76}$ & $0.0479\pm0.0006^{76}$ & $1001^{+66\;76}_{-37}$ & $2.65^{+0.08\;76}_{-0.10}$ & $0.25\pm{0.37}^{77}\tablenotemark{*}$ & $12.7\pm0.8^{77}\tablenotemark{*}$ & Yes\\
        NGTS-5b & $1.136\pm0.023^{78}$ & $0.229\pm0.037^{78}$ & $0.0382\pm0.0013^{78}$ & $952\pm24^{78}$ & $2.643^{+0.066\;78}_{-0.078}$ & $0.19\pm0.11^{77}\tablenotemark{*}$ & $19.5\pm0.5^{77}\tablenotemark{*}$ & Yes\\
        TOI-1807b & $0.122\pm0.008^{80}$ & $0.00809\pm0.00157^{80}$ & $0.0120\pm0.0003^{80}$ & $2100^{+39\;81}_{-40}$ & $3.13\pm2.20$ & $7.58\pm25.52^{79}$ & $118.3\pm14.1^{79}$ & No \\
        TOI-2076b & $0.223\pm0.005^{81}$ & $-$ & $0.0631\pm0.0027^{81}$ & $870\pm13^{81}$ & $-$ & $2.50\pm0.13^{82}$ & $-$ & Yes\\
        TOI-1430.01 & $0.188\pm0.018^{82}$ & $-$ & $0.0705^{82}$ & $813^{82}$ & $-$ & $2.41\pm0.33^{82}$ & $-$ & Yes \\
        TOI-1683.01 & $0.205\pm0.027^{82}$ & $-$ & $0.036^{82}$ & $927^{82}$ & $-$ & $1.94\pm{0.45}^{82}$ & $-$ & Yes \\
        WASP-48b & $1.67\pm0.1^{83}$ & $0.984\pm0.085^{83}$ & $0.0344\pm0.0026^{83}$ & $2035\pm52^{83}$ & $2.91\pm0.06^{83}$ & $0.15\pm0.28$ & $16.3\pm0.4$ & No \\
\enddata
\tablecomments{\scriptsize{Values taken from NASA Exoplanet Archive unless otherwise noted. Error bars are taken from the literature or, if missing, estimated by taking the average error from the remaining data. See page 16 for references.}}
\tablenotetext{*}{\scriptsize{Denotes photometric measurement, which dilutes $\delta_{R_P}$.}}
\end{deluxetable*}
\end{rotatetable*}

\textbf{References for Table \ref{stellar_params}:} 1. \cite{Anderson2017}; 2. \cite{Piaulet2021}; 3. \cite{Kirk2020}; 4. \cite{Anderson2014}; 5. \cite{Nortmann2018}; 6. \cite{Bakos2010}; 7. \cite{Allart2018}; 8. \cite{Bouchy2005}; 9. \cite{Boyajian2015}; 10. \cite{Bonomo2017}; 11. \cite{Barstow2017}; 12. \cite{Salz2018}; 13. \cite{Hebb2009}; 14. \cite{Turner2016};  15. \cite{Kriedberg2018}; 16.\cite{Hartman2011}; 17. \cite{Paragas2021}; 18. \cite{delBurgo2016}; 19. \cite{Melo2006}; 20. \cite{Santos2004}; 21. \cite{Queloz2000}; 22. \cite{Southworth2010}; 23. \cite{Torres2008}; 24. \cite{Alonso2019}; 25. \cite{vonBraun2011}; 26. \cite{Yee2017}; 27. \cite{Bourrier2018}; 28. \cite{Zhang2021}; 29. \cite{Gillon2014}; 30. \cite{Lalitha2014}; 31. \cite{Charbonneau2009}; 32. \cite{Harps2013}; 33. \cite{Kasper2020}; 34. \cite{Niraula2017}; 35. \cite{Rodriguez2018}; 36. \cite{Guo2020}; 37. \cite{Ellis2021}; 38. \cite{Howard2011}; 39. \cite{Salz2015}; 40. \cite{Bourrier2018a}; 41. \cite{Lanotte2014}; 42. \cite{David2019}; 43. \cite{Vissapragada2021}; 44. \cite{Kosiarek2019}; 45. \cite{Bonfils2012}; 46. \cite{Palle2020}; 47. \cite{Gaudi2017}; 48. \cite{Rice2019}; 49. \cite{Prieto2018}; 50. \cite{Carleo2021}; 51. \cite{Triaud2013}; 52. \cite{Gallet2020}; 53. \cite{Triaud2015}; 54. \citep{Fossati2022}; 55. \cite{West2016}; 56. \cite{Casa2021}; 57. \cite{Gillon2016}; 58. \cite{Agol2021}; 59. \cite{Burgasser2017}; 60. \cite{Krishnamurthy2021}; 61. \cite{Gillon2017}; 62. \cite{Mann2020}; 63. \cite{Zhang2022b}; 64. \cite{Hartman2011b}; 65. \cite{Czesla2022}; 66. \cite{Barragan2022}; 67. \cite{Zhang2022c}; 68. \cite{Kawauchi2022}; 69. \cite{Gan2022}; 70. \cite{Hebrard2013};71. \cite{Kirk2022}; 72. \cite{Lam2017}; 73. \cite{dosSantos2020}; 74. \cite{Turner2019}; 75. \cite{Sreejith2020}; 76. \cite{Hartman2011}; 77. \cite{Vissapragada2022b}; 78. \cite{Eigmuller2019}; 79. \cite{Gaidos2023}; 80. \cite{Nardiello2022}; 81. \cite{Hedges2021}; 82. \cite{Zhang2023}; 83. \cite{Enoch2011}; 84. \cite{O'Rourke2014}; 85. \cite{Boro2018}


\begin{acknowledgments}
We would like to thank our reviewer for their very helpful and thoughtful comments. Their insightful improved the quality of this paper. A.O. gratefully acknowledges support from the Dutch Research Council NWO Veni grant.
\end{acknowledgments}

%

\vspace{5mm}
\facilities{Hobby-Eberly Telescope (Habitable-Zone Planet Finder)}


\software{astropy \citep{astropy:2013, astropy:2018, astropy:2022}, barycorrpy \citep{Kanodia2018, Wright2014}, matplotlib \citep{Hunter2007}, numpy \citep{Harris2020}, pandas \citep{McKinney2010}, radvel \citep{Fulton2018}, scipy \citep{Virtanen2020},}




\bibliography{main}{}
\bibliographystyle{aasjournal}



\end{document}